\documentclass[11pt,notitlepage]{article}
\usepackage{fullpage}
\usepackage{epsfig}
\usepackage{graphicx}
\usepackage{amssymb}
\usepackage{amsmath}
\usepackage{multirow}
\usepackage{tabularx}
\usepackage{color}
\usepackage{algorithmic}
\usepackage{algorithm}
\usepackage{caption}
\usepackage{float}
\usepackage{sidecap}
\usepackage{wrapfig}

\usepackage{pgfplotstable}
\usepackage{booktabs}

\newcommand*{\thead}[1]{\multicolumn{1}{|l|}{\bfseries #1}}
\begin{document}

\title{High resolution in-vivo MR-STAT using a matrix-free and parallelized reconstruction algorithm}
\author{Oscar van der Heide$^1$, Alessandro Sbrizzi$^1$, Peter R. Luijten$^1$\\ and Cornelis A. T. van den Berg$^1$}
\date{} 
\maketitle 
\vfill 
% \noindent \textbf{Submitted to Magnetic Resonance in Medicine on 2019-04-29}\\ \\
\textbf{Accepted for publication in NMR in Biomedicine on 2019-12-05}
\\ 
\\
\noindent \textbf{Running title:} High resolution in-vivo MR-STAT\\
\textbf{Word count:} 5777\\
\textbf{Corresponding Author:} Oscar van der Heide, University Medical Center, Heidelberglaan 100, 3508 GA Utrecht, The Netherlands. E-mail: \verb|o.vanderheide@umcutrecht.nl|.\\
 $^1$ Center for Image Sciences, University Medical Center Utrecht, Utrecht, The Netherlands\\
\newpage

\begin{abstract}
  \noindent MR-STAT is a recently proposed framework that allows the reconstruction of multiple quantitative parameter maps from a single short scan by performing spatial localisation and parameter estimation on the time domain data simultaneously, without relying on the FFT. To do this at high-resolution, specialized algorithms are required to solve the underlying large-scale non-linear optimisation problem. We propose a matrix-free and parallelized inexact Gauss-Newton based reconstruction algorithm for this purpose. The proposed algorithm is implemented on a high performance computing cluster and is demonstrated to be able to generate high-resolution ($1mm \times 1mm$ in-plane resolution) quantitative parameter maps in simulation, phantom and in-vivo brain experiments. Reconstructed $T_1$ and $T_2$ values for the gel phantoms are in agreement with results from gold standard measurements and for the in-vivo experiments the quantitative values show good agreement with literature values. In all experiments short pulse sequences with robust Cartesian sampling are used for which conventional MR Fingerprinting reconstructions are shown to fail.\\
  \textbf{Keywords: Quantitative MRI, MR-STAT, MR Fingerprinting, Large-scale non-linear optimization, parallel computing}
  \end{abstract}
 
\newpage

\section*{Abbreviations}
\begin{description}

    \item[bSSFP] Balanced Steady-State Free Precession
    \item[CSF] Cerebrospinal Fluid 
    \item[HFEN] High-Frequency Error Norm 
    \item[L-BFGS] Limited-Memory Broyden-Fletcher-Goldfard-Shanno 
    \item[FFT] Fast Fourier Transform
    \item[MAPE] Mean Absolute Percentage Error 
    \item[MRF] Magnetic Resonance Fingerprinting
    \item[MR-STAT] Magnetic Resonance Spin Tomography in Time-Domain
    \item[NRMSE] Normalized Root Mean Square Error 
    \item[qMRI] Quantitative Magnetic Resonance Imaging
    \item[SVD] Singular Value Decomposition
    \item[TE] Echo time
    \item[TR] Repetition time
    \item[VARPRO] Variable Projection    

\end{description}
\newpage

\section*{Introduction}
\label{sec:introduction}
Conventional magnetic resonance imaging (``MRI") methods rely on the Fourier-Transform relationship between signal and local magnetization value for spatial encoding. Tissue differentiation is possible in the resulting qualitative images because different tissue types have distinct MR-related biophysical properties like $T_1$ and $T_2$ relaxation times. Quantitative MRI (``qMRI") methods aim to estimate MR-related biophysical properties like $T_1$ and $T_2$ relaxation times. Quantitative images could provide additional diagnostic value and are more suited for the purpose of multi-center studies and computer-aided diagnosis \cite{Tofts2003,Deoni2010}. The most straightforward and robust choices for $T_1$ and $T_2$ mapping sequences, i.e. single echo (inversion recovery) spin echo sequences have prohibitively long scan times. Over time, a multitude of alternative pulse sequences have been developed that reduce acquisition times to clinically acceptable levels \cite{Meiboom1958, Look1970, Deoni2003, Deoni2005}. In recent years acquisition times have been reduced even further with advanced reconstruction techniques that include more information of the underlying physical processes in the reconstructions \cite{Ben-Eliezer2015}, add a-priori knowledge in the form of sparsity or low-rank constraints \cite{Zhao2014} and/or allow estimation of multiple parameter maps simultaneously \cite{Teixeira2018,Shcherbakova2018}. A prime example is MR Fingerprinting (``MRF", \cite{Ma2013}). In MRF a transient-state pulse sequence with quasi-random components is used and many highly undersampled k-spaces are acquired in a single, short acquisition. The Fast Fourier Transform (``FFT") is applied on each k-space to generate many snapshot images. Then, on a voxel-per-voxel basis, the measured \emph{fingerprints} are matched to a precomputed Bloch-equation based dictionary to obtain the quantitative parameters. Through this novel combination of transient-state sequences with a pattern recognition step, MRF has been able to drastically reduce qMRI acquisition times.  

MR-STAT \cite{SBRIZZI201856} is a recently proposed qMRI framework that, similarly to MRF, aims to estimate multiple parameter maps from a single short scan simultaneously. However, instead of performing FFTs in a separate step for spatial localisation of signal, parameter maps are fitted directly to the measured time-domain signal using a Bloch-equation based volumetric signal model. That is, a single large-scale non-linear optimisation problem is numerically solved in which the spatial localisation and parameter estimation are performed \emph{simultaneously}. In addition, instead of using a dictionary matching procedure, in MR-STAT gradient-based iterative methods are used to solve the optimisation problem. Compared to MRF the MR-STAT approach results in different trade-offs made in the reconstruction. Since the FFT is no longer explicitly used to transform back and forth between image space and frequency space, the spatial gradient encoding is entangled directly into the MR-STAT signal model. With this approach, data from different readouts of a transient state pulse sequence can be naturally combined into a single reconstruction process. There is no reliance on dictionary compression \cite{Asslander2018} or compressed sensing \cite{Davies2014} techniques to suppress aliasing artefacts. As will be demonstrated, MR-STAT allows for the reconstruction of high quality parameter maps from very short scans even when using standard and experimentally reliable Cartesian sampling strategies.

Solving the non-linear optimisation problem that results from using the volumetric signal model in MR-STAT does introduce new computational challenges. As will be discussed, the computational and memory requirements scale quadratically with the resolution and parallelizing the computations is non-trivial because the FFT is not used to spatially decouple the unknowns. In Sbrizzi et al. \cite{SBRIZZI201856}, to alleviate the computational challenges at high resolution, a 1D FFT along the readout direction was still employed to decouple the problem in one direction in space, resulting in many smaller and independent 1D subproblems to be solved. This hybrid approach only partly benefits from the above mentioned advantages of using a volumetric signal model, e.g., dynamical behaviour during readouts cannot be taken into account. Furthermore, it can only be used with Cartesian sampling strategies. Thirdly, if the technique is applied to 3D acquisitions, each of the resulting 2D subproblems will itself be a large-scale problem. Therefore, to unlock the full potential of MR-STAT, a specialized reconstruction algorithm is required that:
\begin{enumerate}
    \item does not require storage of large model matrices (i.e. is \emph{matrix-free}),
    \item is suitable for a parallel computing implementation to reduce computation times,
    \item is extensible to non-Cartesian sampling strategies.
\end{enumerate}
In the current work we present a reconstruction algorithm for MR-STAT based on an inexact Gauss-Newton method (see \cite{Steihaug1983} and Algorithm 7.2 in \cite{Nocedal1999}) that satisfies the above requirements. For partial derivative computations that are needed in the reconstruction we propose to use exact algorithmic differentiation. With the new reconstruction algorithm we demonstrate the potential of the MR-STAT framework through simulation studies, phantom experiments and by reconstructing high-resolution in-vivo brain maps. Although in principle the reconstruction algorithm can be used with non-Cartesian sampling, in all experiments we will use Cartesian sampling patterns. The reason is that Cartesian sequences - which are used in the vast majority of clinical exams - are challenging to work with in the context of conventional MRF \cite{Stolk2018} whereas with MR-STAT parameter maps can be reconstructed successfully even when using very short acquisitions in the order of seconds per slice.
\section*{Theory}
\label{sec:theory}
In this section we first review the MR-STAT framework as presented by Sbrizzi et al. \cite{SBRIZZI201856}. Then we discuss the computational challenges resulting from the large scale reconstruction problem and we propose techniques to deal with these challenges.

\subsection*{MR-STAT Framework}

    The time evolution of a single spin isochromat $\mathbf{m}=\left( m_x, m_y, m_z \right)$ with spatial coordinates $\mathbf{r} = (x,y,z)$ and tissue properties $\boldsymbol{\theta} = \left(T_1, T_2, \ldots \right)$ is governed by the Bloch equations.

    Let $m = m_x + i m_y$ be the transverse component of the magnetization in the rotating frame. The demodulated time-domain signal $s$ is equal to the volume integral of the transverse magnetisation of all spins within the field of view $V$, weighted by their effective proton spin densities $\rho$. For the purpose of this work, $\rho$ includes also the amplitude of the receive coil sensitivity and the transceive phase, thus $\rho$ is a complex quantity, i.e. $\rho = \rho_x + i \rho_y$. In short:

    \begin{align}
        s(t)=\int_{V}\rho(\boldsymbol{r})m(\boldsymbol{r},\boldsymbol{\theta}(\boldsymbol{r}),t)d\boldsymbol{r}.\label{eq:1}
    \end{align}
    After discretization of the field of view $V$ into $N_v$ voxels, each having volume $\Delta_V$, equation \eqref{eq:1} becomes
    \begin{equation}
        s(t)=\sum_{j=1}^{N_v} \rho_{j}m_{j}(\boldsymbol{\theta}_j,t)\Delta_V.
    \end{equation}
    Here $m_j$ is the magnetization in voxel $j$, which can be computed by numerical integration of the Bloch equations.

    Let $N_t$ be the total number of signal samples and let $t_1, \ldots, t_{N_t}$ denote the sampling times. Define the magnetization vector $\mathbf{m}_j$ in voxel $j$ as
    \begin{equation}
        \mathbf{m}_j := \left( m_j(\boldsymbol{\theta}_j,t_1), \ldots, m_j(\boldsymbol{\theta}_j,t_{N_t}) \right)\in\mathbb{C}^N_t
    \end{equation}
    and the signal vector $\mathbf{s} \in \mathbb{C}^{N_t}$ as
    \begin{equation}
        \textbf{s} = \sum_{j=1}^{N_v} \rho_{j}\mathbf{m}_{j}
    \end{equation}
    Note that if we introduce the \emph{magnetization matrix} $\textbf{M} \in \mathbb{C}^{N_t \times N_v}$,
    \begin{equation}
        M_{i,j} := m_j(\boldsymbol{\theta}_j, t_i),
    \end{equation}
    and proton density vector $\boldsymbol{\rho} \in \mathbb{C}^{N_v}$
    \begin{equation}
    \boldsymbol{\rho} = \left( \rho_1, \ldots, \rho_{N_v} \right),
    \end{equation}
    then $\textbf{s}$ can be written as
    \begin{equation}
        \mathbf{s} = \mathbf{M}\boldsymbol{\rho}.\label{eq:rholin}
    \end{equation}

    Let $N_p$ denote the number of distinct parameters per voxel (including real and imaginary parts of the proton density). Then ${s}$ depends on $N:= N_v \times N_p$ different parameters. All parameters are concatenated into a single vector $\boldsymbol{\alpha} \in \mathbb{R}^{N}$ in such a way that indices $\left\{j + kN_v  \mid k=0\ldots, N_p-1 \right\} $ denote the parameters associated with voxel $j$.

    Now, given a vector of measured time-domain samples $\mathbf{d} \in \mathbb{C}^{N_s}$, define the residual vector $\textbf{r}\in \mathbb{C}^{N_t}$ as
    \begin{align}
        \mathbf{r}(\boldsymbol{\alpha}) = \mathbf{d}-\mathbf{s}(\boldsymbol{\alpha})
    \end{align}
    and define the non-linear least-squares objective function $f:\mathbb{R}^N \rightarrow \mathbb{R}$ as
    \begin{align}
        f(\boldsymbol{\alpha})=\frac{1}{2}\|\mathbf{r}(\boldsymbol{\alpha})\|_{2}^{2}
    \end{align}

    The parameter maps $\boldsymbol{\alpha}^{*}$ are obtained by numerically solving
    \begin{equation}
        \boldsymbol{\alpha}^{*} = \text{argmin}_{\boldsymbol{\alpha}}f(\boldsymbol{\alpha}) \label{eq:2},
    \end{equation}
    subject to physical constraints represented by the Bloch equations and realistically attainable intervals for the parameters.

    \subsection*{Computational Challenges}

    Note that \eqref{eq:2} is a non-linear optimization problem that requires iterative algorithms to be solved. At each iteration, the signal $\mathbf{s} = \mathbf{M}\boldsymbol{\rho}$ needs to be computed and that requires the Bloch equations to be integrated for each voxel. In addition, the gradient of $f$ (i.e. the vector of partial derivatives of $f$ with respect to each of the parameters) needs to be computed. From the least-squares structure of the problem it follows that the gradient can be expressed as
    \begin{equation}
        \mathbf{g} \:= \nabla f = \Re(\mathbf{J}^H \mathbf{r}),
    \end{equation}
    where $\mathbf{J} \in \mathbb{C}^{N_t \times N_v}$ is the Jacobian matrix defined as
    \begin{equation}
        \mathbf{J}(\boldsymbol{\alpha}):=\left[ \frac{\partial \mathbf{r}}{\partial {\alpha}_1} \ldots \frac{\partial \mathbf{r}}{\partial {\alpha}_{N}}\right],
    \end{equation}
    $\mathbf{J}^H$ is the Hermitian transpose of $\mathbf{J}$ and $\Re$ is the real-part operator.

    A gradient-descent type algorithm could be used to minimize \eqref{eq:2} but it may result in poor convergence (see Chapter 3 of Nocedal and Wright \cite{Nocedal1999}). Second-order methods (i.e. Newton methods) typically lead to better convergence. At each iteration, these methods require the inversion of a linear system involving (an approximation to) the Hessian matrix $\mathbf{H} \in \mathbb{R}^{N \times N}$, which includes curvature information and is defined as
    \begin{equation}
        \mathbf{H}(\boldsymbol{\alpha}):=\left[\frac{\partial^{2}f}{\partial\alpha_i\partial \alpha_j}\right]_{i,j=1}^{N_p}.
    \end{equation}
    A second-order MR-STAT reconstruction algorithm would follow the steps as outlined in Algorithm \ref{alg:alg1}:

    % \begin{algorithm}
    %     \caption{Minimize $f(\boldsymbol{\alpha}) = \frac{1}{2}\|\mathbf{d} - \mathbf{s(\boldsymbol{\alpha})} \|_2^2$}
    %     \begin{algorithmic}[0]
    %         \REQUIRE Initial guess $\boldsymbol{\alpha}$
    %         \WHILE{not converged}
    %         \STATE 1. Compute residual: $\mathbf{r} = \mathbf{d} - \mathbf{s} = \mathbf{d} - \mathbf{M}\boldsymbol{\rho}$
    %         \STATE 2. Compute gradient: $\mathbf{g} = \Re \left( \mathbf{J}^H \mathbf{r} \right)$
    %         \STATE 3. Solve linear system: $\mathbf{H} \mathbf{p} = -\mathbf{g}$
    %         \STATE 4. Update parameters: $\boldsymbol{\alpha} = \boldsymbol{\alpha} + \mathbf{p}$
    %         \ENDWHILE
    %     \end{algorithmic}
    %     \label{alg:alg1}
    % \end{algorithm}

    Using Algorithm \ref{alg:alg1} for MR-STAT poses several practical challenges due to the large scale of the problem.

    First of all, to estimate $N$ parameters, the number of sample points $N_t$ will in general be in the order of $N=N_p \times N_v$ as well. Assuming $N_t \approx N$, it follows that $\mathbf{M}$ will be of size $N_t \times N_v \approx (N_pN_v) \times N_v$ (complex entries) and $\mathbf{J}$ will be of size $2N_t \times N \approx (N_pN_v) \times (N_pN_v)$ (complex entries). Since $\mathbf{H}$ will be of size $(N_pN_v) \times (N_pN_v)$ as well, it follows that all three matrices scale with $N_v^2$. In Table \ref{tab:matrixsizes}, the required computer memory to store matrices of these sizes is reported for various values of $N_v$ for the case $N_p=4$. It can be seen that, even for 2D acquisitions, it will be infeasible to store these matrices in memory for clinically relevant resolutions.
    
    Secondly, the actual time needed to compute the entries of $\mathbf{M}$, $\mathbf{J}$ and $\mathbf{H}$ scales with $N_v^2$ as well. When using a regular desktop computer the reconstruction times quickly become too long to make MR-STAT useful in clinical practice.

    Fortunately, as will be detailed in the next section, Algorithm \ref{alg:alg1} only requires matrix-vector products with the matrices $\mathbf{M}, \mathbf{J}$ and (approximations to) $\mathbf{H}$. These matrix-vector product can be computed without having to store the full matrices in memory. Moreover, the computation of the matrix-vector products can be efficiently distributed among multiple computing cores on a high performance computing cluster, reducing the MR-STAT computation times to acceptable levels for off-line reconstructions.

\subsection*{Solution Strategies}

    \subsubsection*{Computing the time-domain signal $\textbf{s}$}

    In the first step of Algorithm \ref{alg:alg1} we need to compute $\mathbf{r} = \mathbf{d} - \mathbf{s}$ for the current estimate of the parameters $\boldsymbol{\alpha}$. Recall that
    \begin{equation}
        \mathbf{s} = \mathbf{M}\boldsymbol{\rho} = \sum_{j=1}^{N_v} \rho_j \mathbf{m}_j.
    \end{equation}
    Since the time evolution of the magnetization in each voxel is assumed to be independent from other voxels, the $ \mathbf{m}_j $ can be computed independently from each other. In particular, storage of the matrix $\mathbf{M}$ is not required for computing $\mathbf{s}$, see Algorithm \ref{alg:algMv}.

    % \begin{algorithm}
    %     \caption{Compute $\mathbf{s}(\boldsymbol{\alpha}) = \mathbf{M}\boldsymbol{\rho}$ (matrix-free, serial)}
    %     \begin{algorithmic}[0]
    %         \STATE Initialize $\mathbf{s}$ = zeros($N_t$,1)
    %         \FOR{j $\leftarrow$ $1$ to $N_v$}
    %         \STATE 1. Integrate Bloch equations in time to obtain $\mathbf{m}_j = \left[ m_j(t_1), \ldots, m_j(t_{N_t})\right]^T$
    %         \STATE 2. Set $\mathbf{s} = \mathbf{s} + \rho_j \mathbf{m}_j$
    %         \ENDFOR
    %         \STATE Return $\mathbf{s}$
    %     \end{algorithmic}
    %     \label{alg:algMv}
    % \end{algorithm}

    Note that Algorithm \ref{alg:algMv} only requires the allocation of two vectors of length $N_t$, which is feasible on modern computing architectures for both 2D and 3D acquisitions. The computation of $\mathbf{s}$ can then be parallelized using $N_c$ computing cores by following the procedure outlined in Algorithm \ref{alg:algMv_par} (see also \cite{Stocker2010,Liu2017}).
    
    % \begin{algorithm}
    %     \caption{Compute $\mathbf{s}(\boldsymbol{\alpha}) = \mathbf{M}\boldsymbol{\rho}$ (matrix-free, parallel)}
    %     \begin{algorithmic}[0]
    %         \REQUIRE Master process $p_m$, slave processes $p_i$ for $i \in [1, \ldots, N_c]$.
    %         % \STATE Initialize $\mathbf{s}$ = zeros(Ns,1)
    %         \STATE 1. $p_m$ distributes $\boldsymbol{\alpha}$: each $p_i$ receives the parameters associated with voxels $ [(i-1) * N_v + 1, \ldots, i*N_v]$.
    %         \STATE 2. Each $p_i$ uses Algorithm \ref{alg:algMv} to compute a  ``local'' version of the signal $\tilde{\mathbf{s}}_{i}$.
    %         \STATE 3. Each $p_i$ communicates $\tilde{\mathbf{s}}_{i}$ back to the $p_m$.
    %         \STATE 4. On $p_m$ the signal $\mathbf{s}$ is computed as $\mathbf{s} = \sum_{i=1}^{N_c} \tilde{\mathbf{s}}_{i}$.
    %     \end{algorithmic}
    %     \label{alg:algMv_par}
    % \end{algorithm}

    The communication requirements for this parallelized algorithm can be summarized as follows:
    \begin{itemize}
    \item To distribute the parameters, the master process sends $N / N_c$ parameters to each of the $N_c$ slaves.
    \item To receive the local signals from the slaves, each slave sends a vector of length $N_t \approx N $ to the master process.
    \end{itemize}

    \subsubsection*{Computing the gradient $\textbf{g}$}

    To compute $\textbf{g} = \nabla f$ for the current estimate of the parameters $\boldsymbol{\alpha}$, recall that
    \begin{equation}
        \textbf{g} = \Re(\textbf{J}^H\textbf{r}).
    \end{equation}
    Since $\mathbf{J}$ is defined as
    \begin{equation}
        \mathbf{J} = \left[ \frac{\partial \mathbf{r}}{\partial {\alpha}_1} \ldots \frac{\partial \mathbf{r}}{\partial {\alpha}_{N}}\right],
    \end{equation}
    it follows that
    \begin{equation}
        \textbf{g} =
        \begin{bmatrix}
            \Re \left( \left\langle \frac{\partial \mathbf{r}}{\partial {\alpha}_1} , \mathbf{r} \right\rangle \right) \\
            \vdots \\
            \Re \left( \left\langle \frac{\partial \mathbf{r}}{\partial {\alpha}_N}, \mathbf{r} \right\rangle \right) \\
        \end{bmatrix}.
    \end{equation}

    To compute the $\frac{\partial \textbf{r}}{\partial \alpha_{i}}$, again note that the magnetization in different voxels is assumed to evolve independently. Hence if $\alpha_i$ is a parameter associated with voxel $j$ (i.e. $j = i \mod N_v$), it follows that
    \begin{equation}
        \frac{\partial \mathbf{r}}{\partial \alpha_{i}} =
        - \frac{\partial \left( \sum_{r=1}^{N_v} \rho_r \mathbf{m}_{r} \right) }{\partial \alpha_{i}} =
        - \frac{\partial \left( \rho_j \mathbf{m}_{j} \right) }{\partial \alpha_{i}}.
    \end{equation}

    % \begin{algorithm}
    %     \caption{Compute $\mathbf{g}(\boldsymbol{\alpha}) = \Re(\mathbf{J}^H \mathbf{r})$ (matrix-free, serial)}
    %     \begin{algorithmic}[0]
    %         \STATE Initialize $\mathbf{g}$ = zeros($N$,1)
    %         \FOR{j $\leftarrow$ 1 to $N_v$}
    %         \FOR{k $\leftarrow$ 1 to $N_p$}
    %         \STATE 1. Set $i = j + (k-1)N_v$
    %         \STATE 2. Compute partial derivative $\frac{\partial \mathbf{r}}{\partial \alpha_{i}}  = -\frac{\partial \left(\rho_j \mathbf{m}_{j} \right)}{\partial \alpha_{i}}$
    %         \STATE 3. Set $\mathbf{g}[i] = \Re \left( \left<\frac{\partial \mathbf{r}}{\partial \alpha_{i}}, \mathbf{r} \right> \right)$
    %         \ENDFOR
    %         \ENDFOR
    %         \STATE Return $\mathbf{g}$
    %     \end{algorithmic}
    %     \label{alg:algJtv}
    % \end{algorithm}

    Using Algorithm \ref{alg:algJtv} only requires storage of one vector of length $N$ for the output and - in principle - one complex vector of length $N_t$ to store the intermediate partial derivative vector. In practice we will compute the $N_p$ partial derivatives for each voxel simultaneously so that $N_p$ complex vectors of length $N_t$ are stored simultaneously. For a high-resolution 2D scan this only requires limited memory (in the order of tens of megabytes).

\begin{figure}
\centering
\includegraphics[width=.99\textwidth]{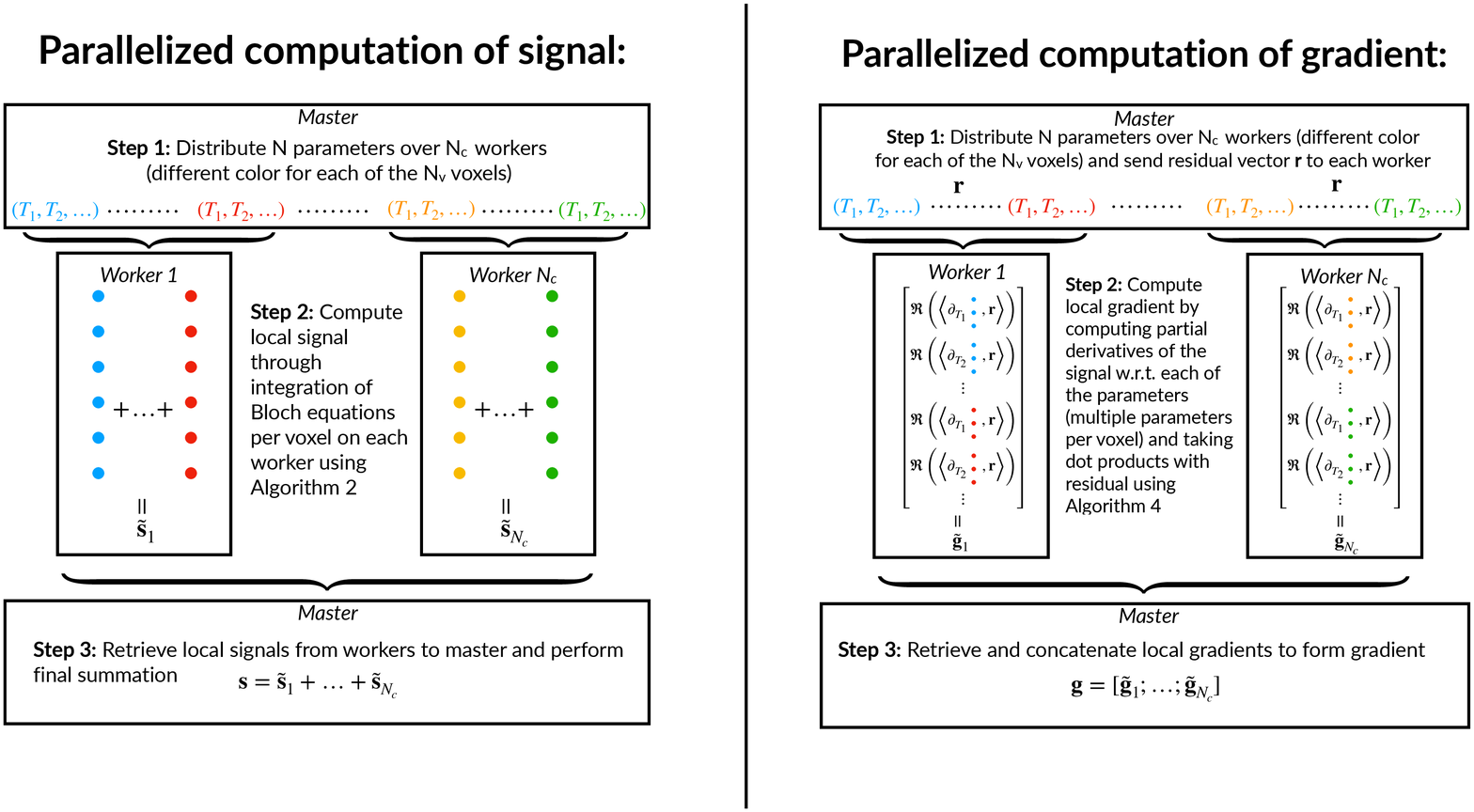}
\caption{Visualization of the (matrix-free) algorithms to compute the signal [left] and the gradient [right] in a parallelized fashion.}\label{fig:paralgs}
\end{figure}

    Utilizing Algorithm \ref{alg:algJtv}, the computation of $\mathbf{g}$ can be performed in parallel as outlined in Algorithm \ref{alg:algJtv_par}. The parallelization schemes for both the signal and gradient computations are visualized in Figure \ref{fig:paralgs}.

    % \begin{algorithm}
    %     \caption{Compute $\mathbf{g}(\boldsymbol{\alpha}) = \Re (\mathbf{J}^H \mathbf{r})$ (matrix-free, parallel)}
    %     \begin{algorithmic}[0]
    %         \REQUIRE Master process $p_M$, slave processes $p_i$ for $i \in [1, \ldots, N_c]$.
    %         \STATE 1. $p_m$ distributes $\boldsymbol{\alpha}$: each $p_i$ receives the parameters associated with voxels $ [(i-1) * N_v + 1, \ldots, i*N_v]$.
    %         \STATE 2. $p_m$ distributes $\mathbf{r}$ to each $p_i$.
    %         \STATE 3. Each $p_i$ uses Algorithm \ref{alg:algJtv} to compute a  ``local'' gradient $\tilde{\mathbf{g}}_{i}$.
    %         \STATE 4. Each $p_i$ communicates $\tilde{\mathbf{g}}_{i}$ back to the $p_m$.
    %         \STATE 5. On $p_m$ the gradient $\mathbf{s}$ is computed by vertical concatenation of the $\tilde{\mathbf{g}}_{i}$.
    %     \end{algorithmic}
    %     \label{alg:algJtv_par}
    % \end{algorithm}
    
    Communication requirements for the parallel gradient computation can be summarized as follows:
    \begin{itemize}
        \item To distribute the parameters, the master process sends $N / N_c$ parameters to each of the $N_c$ slaves.
    \item To distribute the residual vector the master process sends a vector of length $N_t$ to each slave.
    \item To receive the local gradients from the slaves, each slave sends a vector of length $N / N_c$ to the master process.
    \end{itemize}

\begin{figure}
    \centering
    \includegraphics[width=.75\textwidth]{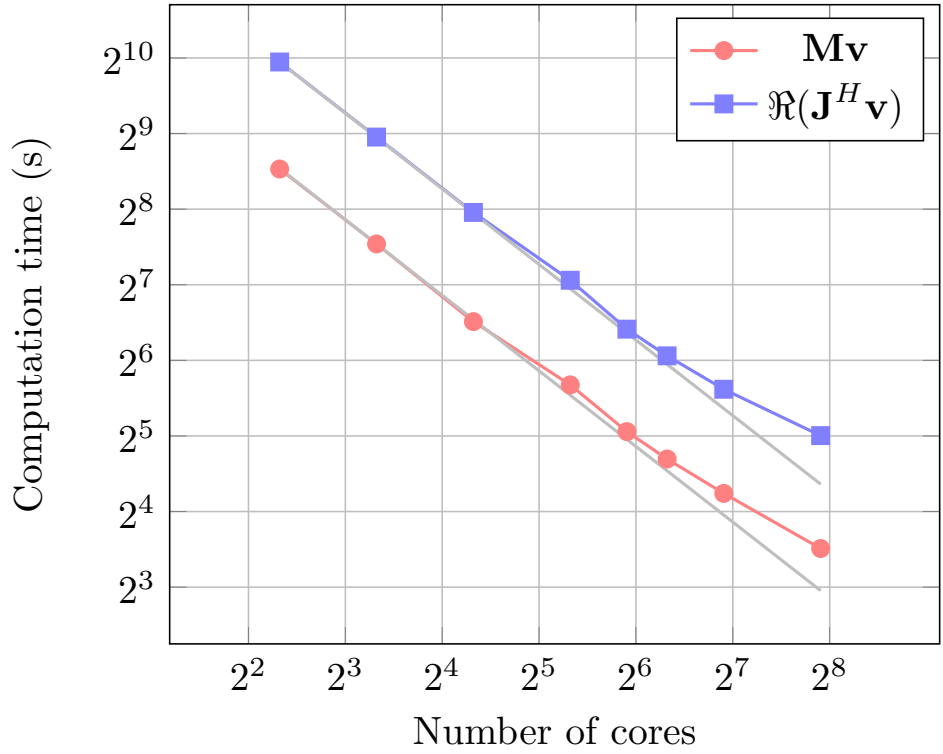}
    \caption{Time needed to compute matrix-vector products of the form $\mathbf{M}\mathbf{v}$ and $\mathbf{J}^H\mathbf{v}$ for different numbers of cores $N_c$ used on a high performance computing cluster.}
    \label{fig:invivo_parallelization}
\end{figure}

    Note that for both algorithms \ref{alg:algMv_par} and \ref{alg:algJtv_par}, the communication requirements scale linearly with the number of parameters $N$ for a fixed number of cores $N_c$. Since $N = N_v \times N_p$, it follows that the communication requirements scale linearly with $N_v$ as well. As discussed in the previous section the computational requirements scale quadratically with $N_v$. Therefore we hypothesize that, as long as $N_c \ll N_v$, the communication overhead is negligible compared to reduction in computation times achieved by dividing the computation load over $N_c$ computing cores. That is, we expect the total computation time to decrease linearly with the number of cores available under this assumption. This hypothesis is confirmed in Figure \ref{fig:invivo_parallelization} in the results section.

\subsubsection*{Incorporating Curvature Information}

Given the ability to compute the gradient $\mathbf{g}$ using the matrix-free, parallelized algorithm from the previous subsection, in principle the so called limited-memory Broyden--Fletcher--Goldfarb--Shanno (``L-BFGS", \cite{Nocedal1980}) method can be applied to obtain the update step $\mathbf{p}$ at each iteration. The L-BFGS method approximates the inverse of the Hessian matrix using a limited number of gradient vectors from previous iterations. However, in practice the L-BFGS method was observed to result in poor convergence. 

Alternatively, since we are dealing with a least-squares problem, a Gauss-Newton method might be used in which the Hessian matrix $\mathbf{H}$ in Algorithm \ref{alg:alg1} is approximated by $\Re (\mathbf{J}^H\mathbf{J})$ and
\begin{equation}
    \Re \left(\mathbf{J}^H\mathbf{J}\right) \mathbf{p} = -\mathbf{g} \label{eq:4}
\end{equation} is solved to obtain update steps $\mathbf{p}$. Note that the matrix $\Re ( \mathbf{J}^H\mathbf{J} )$ is of the same size as the Hessian matrix itself and thus, in principle, cannot be stored into computer memory. If, however, we use iterative techniques (e.g. a Conjugate Gradient method) to solve the linear system $\Re (\mathbf{J}^H\mathbf{J}) \mathbf{p} = -\Re (\mathbf{g})$, we only need matrix-vector products with $\Re (\mathbf{J}^H\mathbf{J})$. In the previous subsection it was outlined how matrix-vector products of the form $\mathbf{J}^H\mathbf{v}$ may be computed in a matrix-free, parallelized fashion. Similar techniques can be applied to matrix-vector products of the form $\mathbf{J}\mathbf{v}$. Hence matrix-vector products of the form $\Re (\mathbf{J}^H\mathbf{J})\mathbf{v}$ can be computed in a matrix-free, parallelized fashion by first computing $\mathbf{y} = \mathbf{J}\mathbf{v}$ and subsequently computing $\Re(\mathbf{J}^H \mathbf{y})$. With this technique, the linear system in equation (\ref{eq:4}) can be solved numerically even for large scale problems. In practice it will not be necessary to solve equation (\ref{eq:4}) to high precision and the number of iterations in this \emph{inner loop} can be limited, resulting in an \emph{inexact} Gauss-Newton method (see \cite{Steihaug1983} and Algorithm 7.2 in \cite{Nocedal1999}) as outlined in Algorithm \ref{alg:inexact_gn}

    % \begin{algorithm}
    %     \caption{(Inexact Gauss-Newton MR-STAT) Minimize $f(\boldsymbol{\alpha}) = \frac{1}{2}\|\mathbf{d} - \mathbf{s(\boldsymbol{\alpha})} \|_2^2$}
    %     \begin{algorithmic}
    %         \REQUIRE Initial guess $\boldsymbol{\alpha_0}$, initial trust radius $\Delta_0$
    %         \STATE
    %         \WHILE{not converged}
    %         \STATE Compute $\mathbf{s}$ (matrix-free, parallel)
    %         \STATE Set $\mathbf{r} = \mathbf{d} - \mathbf{s}$
    %         \STATE Compute $\mathbf{g} = \mathbf{J}^H \mathbf{r}$ (matrix-free, parallel)
    %         \STATE Solve with CG iterations (inner GN loop):
    %         \STATE $\quad \Re \left(\mathbf{J}^H \mathbf{J} \right) \mathbf{p} = -\mathbf{g}$ (matrix-free, parallel)
    %         \STATE Set $\boldsymbol{\alpha} = \boldsymbol{\alpha} + \mathbf{p}$
    %         \ENDWHILE
    %     \end{algorithmic}
    %     \label{alg:inexact_gn}
    % \end{algorithm}

    \section*{Methods}
    \label{sec:methods}
    The matrix-free, parallelized MR-STAT reconstruction algorithms was tested on both simulated and experimentally acquired data.
    
    \subsection*{Pulse Sequence}
    
    In all test cases, a transient-state 2D balanced gradient-echo pulse sequence similar to the pulse sequence in Sbrizzi et al \cite{SBRIZZI201856} was used. Throughout the whole sequence the TR was fixed and TE was set to TR/2. A linear, Cartesian sampling strategy was employed together with time-varying flip angles that change according to a smoothly varying pattern. We refer to Supporting Information S1 for more details on the sampling trajectory and flip angle pattern. The phase of the RF pulse alternated between $0$ and $180$ degrees. Changing the flip angles prevents the spins from reaching a steady-state and by following a smoothly varying pattern the spin-echo behaviour of bSSFP sequences \cite{Scheffler2003} is preserved to a large extent. This spin-echo like behaviour is needed for proper $T_2$ estimation and at the same time it also effectively eliminates sensitivity to $\Delta B_0$ within a certain passband of off-resonances \cite{Asslander2017}. An added benefit of the smoothly changing flip angle train is the improved convexity of the minimization landscape \cite{Sbrizzi2017}.
    
    Each RF pulse has a Gaussian envelope and at the start of the pulse sequence a non-selective inversion pulse is played out for enhanced $T_1$ encoding. The pulse sequence was implemented on a 1.5 T clinical MR system (Ingenia, Philips Healthcare, Best, The Netherlands).
    
    \subsection*{Reconstructions}
    
    All reconstruction code was written in the open-source Julia programming language \cite{DBLP:journals/corr/BezansonEKS14}. To compute the MR-signal for a given set of parameters, an optimized Bloch-equation solver was implemented which also takes into account also the slice profile \cite{Valenberg2015}. To compute exact partial derivatives algorithmic differentiation in forward mode \cite{Wengert2002} was implemented. We refer to the Supporting Information S2 for more details.
    
    The inexact Gauss-Newton method was implemented using a trust-region framework (following \cite{Steihaug1983} and Algorithm 7.2 in \cite{Nocedal1999}). In order to facilitate bound constraints on the parameters, reflection at feasible boundaries was incorporated \cite{Coleman1996}. For the L-BFGS method, an implementation from the Optim.jl package \cite{KMogensen2018} was used. The reconstruction algorithm was implemented on a high performance computing cluster which consists of multiple Intel Xeon Gold 6148 nodes with 40 cores each, on which the CentOS Linux 7 (Core) operating system is installed.
    
    For all experiments, $T_1,T_2$ and $\rho$ (complex) maps are reconstructed. For the data obtained with clinical MR systems we also reconstruct $|B_1^+|$ to take into account transmit field inhomogeneities. The off-resonance $\Delta B_0$ was set to zero and thus it was not reconstructed because of the flat spectral response of the balanced sequence within the passband. The non-linear parameters were initialized as follows: $T_1 = 1000$ ms, $T_2 = 100$ ms, $|B_1^+| = 1\,$ a.u. and $\Delta B_0 = 0$ Hz. In previous work \cite{SBRIZZI201856} the Variable Projection method (``VARPRO", \cite{Golub2003}) was utilized to separate the linear parameters (i.e. proton density) from the non-linear parameters. The VARPRO method in principle requires computing (through SVD or QR decomposition) and storing an orthogonal basis for the matrix $\mathbf{M}$. For the matrix sizes in the current work that would be computationally infeasible and it is non-trivial to extend the VARPRO technique to matrix-free methods. Therefore, in the current work we treat the proton densities as non-linear parameters. We only make use of the linearity to provide an initial guess for the proton densities. That is, given the initial guess for the non-linear parameters, the (complex) proton density was initialized as the least squares solution to the linear system $\mathbf{M}(\boldsymbol{\alpha}_0)\boldsymbol{\rho} = \mathbf{d}$ obtained using a linear solver (LSQR). Based on the resulting initial guess for the proton density, a mask was drawn to exclude regions with no significant proton density from subsequent simulations.
    
    In all reconstructions, logarithmic scaling is applied to both $T_1$ and $T_2$ parameters. The variable substitution brings both variables in a similar range and it thus improves convergence of the algorithm.

    The reconstruction code will be made available online at https://gitlab.com/mrutrecht/mrstat after acceptance of the manuscript for publication.
    
    \subsection*{Numerical Brain Simulation}
    Signal from a numerical brain phantom \cite{Aubert-Broche2006} with a field-of-view of $192$ mm $\times$ $192$ mm and voxel size of $1$ mm $\times$ $1$ mm was simulated using the transient-state pulse sequence. A total number of $8\times192 = 1536$ readouts were simulated (each phase encoding line was acquired eight times but note that for each readout line the flip angle and thus state of the spins is different) with a TR of $7.88$ ms and a TE of $3.94$ ms. The total sequence duration was $12.1$ s.

    Reconstructions were performed using 64 cores. The number of outer and inner iterations for the inexact Gauss-Newton method were limited to 40 and 20, respectively.

    For comparison purposes, we also performed MRF reconstructions on signal from the numerical brain phantom using the Cartesian trajectory, as well as signal from radial and spiral trajectories for which MRF is known to work well. In all three cases the same flip angle train, TE and TR were used. For the radial case, $k_{max}$ was extended by a factor of $√2$ and each readout the spoke was rotated by the golden angle. For the spiral acquisition a variable density spiral was generated \cite{Lee2003,Jiang2015} that would require 24 interleaves to fully sample the inner region of k-space and 48 interleaves for the outer region of k-space. The spiral was rotated by the golden angle each readout. Data $\mathbf{d}^{MRF}$ was then simulated by applying a forward operator, consisting of the (non-uniform) FFT \cite{Fessler2003} and an undersampling operator, on fingerprints simulated using the numerical brain phantom. To perform the MRF reconstructions, a dictionary consisting of 100 $T_1$ values from $0.1$ s to $5.0$ s in increments of $4$ \% and 100 $T_2$ values from $0.01$ s to $2.0$ s in increments of $5.5$ \% was generated. The $T_1$ and $T_2$ values of the phantom were not used in generating the dictionary. The dictionary was compressed in the time direction to rank 5 \cite{McGivney2014} using the SVD. For all trajectories (linear) low-rank forward operators $\boldsymbol{E}^{MRF}$ were formed that consisted of the expansion of low-rank coefficients to the full time-series, a nuFFT operator, and a sampling operator compression \cite{Asslander2018}. Low-rank snapshot images $\mathbf{x}^{MRF}$ were reconstructed from the undersampled data $\mathbf{d}^{MRF}$ by solving the linear system
    \begin{equation}
        \mathbf{x}^{MRF} = \text{argmin}_{\mathbf{x}} \| \mathbf{d}^{MRF} - \boldsymbol{E}^{MRF}\mathbf{x} \|_{2}^{2}
    \end{equation}
    with LSQR (similar to e.g. \cite{Zhao2018a} and low-rank inversion in \cite{Asslander2018}). Finally, dictionary matching with the compressed dictionary was performed on $\mathbf{x}^{MRF}$ to obtain the parameter estimates.

    To further study the effect of noise on the MR-STAT reconstruction algorithm, additional reconstructions were performed where complex Gaussian noise was added to the simulated signal such that $\|\text{signal}\|_{2} / \| \text{noise}\|_{2} = 50, 25$ and $10$.
    
    \subsection*{Gel Phantom Experiment}
    
    Signal from a 2D transverse slice of six gadolinium-doped gel phantoms (TO5, Eurospin II test system, Scotland) was collected on the 1.5 T MR system using the manufacturer's thirteen-channel receive headcoil. In total  $8\times96 = 768$ readout lines were acquired with a spatial resolution of $1$ mm $\times$ $1$ mm $\times$ 5 mm and a field-of-view of $96$ mm $\times$ $96$ mm. The TR and TE were $7.4$ ms and $3.7$ ms, respectively, resulting in a total acquisition time of $5.7$ s. For reproducibility purposes the MR-STAT scan was repeated four times with full relaxation in between the different scans. 
    
    Parameters that describe the pulse sequence were exported from the scanner and subsequently loaded into Matlab \cite{MATLAB2015}. The measured signals from different receive channels were compressed into a single signal by applying the principal component analysis and choosing the principle mode \cite{Buehrer2007}.
    
    Reconstructions of the parameter maps were performed using the inexact Gauss-Newton method on the computing cluster using 32 cores. The number of inner iterations was limited to fifteen whereas the number of outer iterations was limited to ten.
    
    To assess correctness of the $T_1$ and $T_2$ maps reconstructed with MR-STAT, data was also acquired using gold standard methods in the form of an inversion-recovery single spin-echo protocol with inversion times of [50, 100, 150, 350, 550, 850, 1250] ms for $T_1$ mapping and a single echo spin-echo protocol with echo times of [8, 28, 48, 88, 138, 188] ms for $T_2$ mapping. 

    \subsection*{In-vivo experiments}
    
    Using the 1.5 T clinical MR system we also acquired signal from 2D transverse slices of the brain in three healthy volunteers. Each volunteer gave written informed consent. For each acquisition, a total of  $8\times192 = 1536$ readout lines were acquired with acquisition parameters as reported in Table \ref{tab:acqpars}. After masking approximately 25000 voxels remain for which quantitative parameters are estimated. The MR-STAT reconstructions were performed with 64 cores using the reconstruction settings as for the gel phantom experiment.
    
    To demonstrate the effect of accelerated acquisitions, we also performed reconstructions using time-domain data corresponding to the first 896 TRs and the first 448 TRs from one of the subjects. The corresponding acquisition times were $6.8$ s and $3.4$ s respectively.
    
    One of the in-vivo brain datasets was also used to test the effectiveness of the parallelization scheme. Individual matrix-vector products of the form $\mathbf{M}\mathbf{v}$ and $\mathbf{J}^H\mathbf{v}$ were computed and timed for $5, 10, 20, 40, 60, 80, 100, 120$ and $240$ cores respectively.
    
\section*{Results}

\label{sec:results}
\subsection*{Parallelization}

In Figure \ref{fig:invivo_parallelization} the time required to compute matrix-vector products of the form $\mathbf{M}\mathbf{v}$ and $\mathbf{J}^H\mathbf{v}$ for the one of the in-vivo datasets is shown for an increasing number of computing cores $N_c$. Initially we observe a linear decrease in computation times, however this linear decrease flattens beyond approximately 64 cores. This effect can be explained by the increase in communication overhead when using more cores and increased competition between cores for shared resources like memory bandwidth and cache memory. Although the linear decrease flattens beyond 64 cores, a decrease in computation times is still observed even when going towards 240 cores. Because for MR-STAT reconstruction times are dominated by the computation of these matrix-vector products, the reconstruction times can thus be effectively reduced by the proposed parallel implementation.

\subsection*{Numerical Brain Phantom}

\begin{figure}
    \centering
    \includegraphics[width=.99\textwidth]{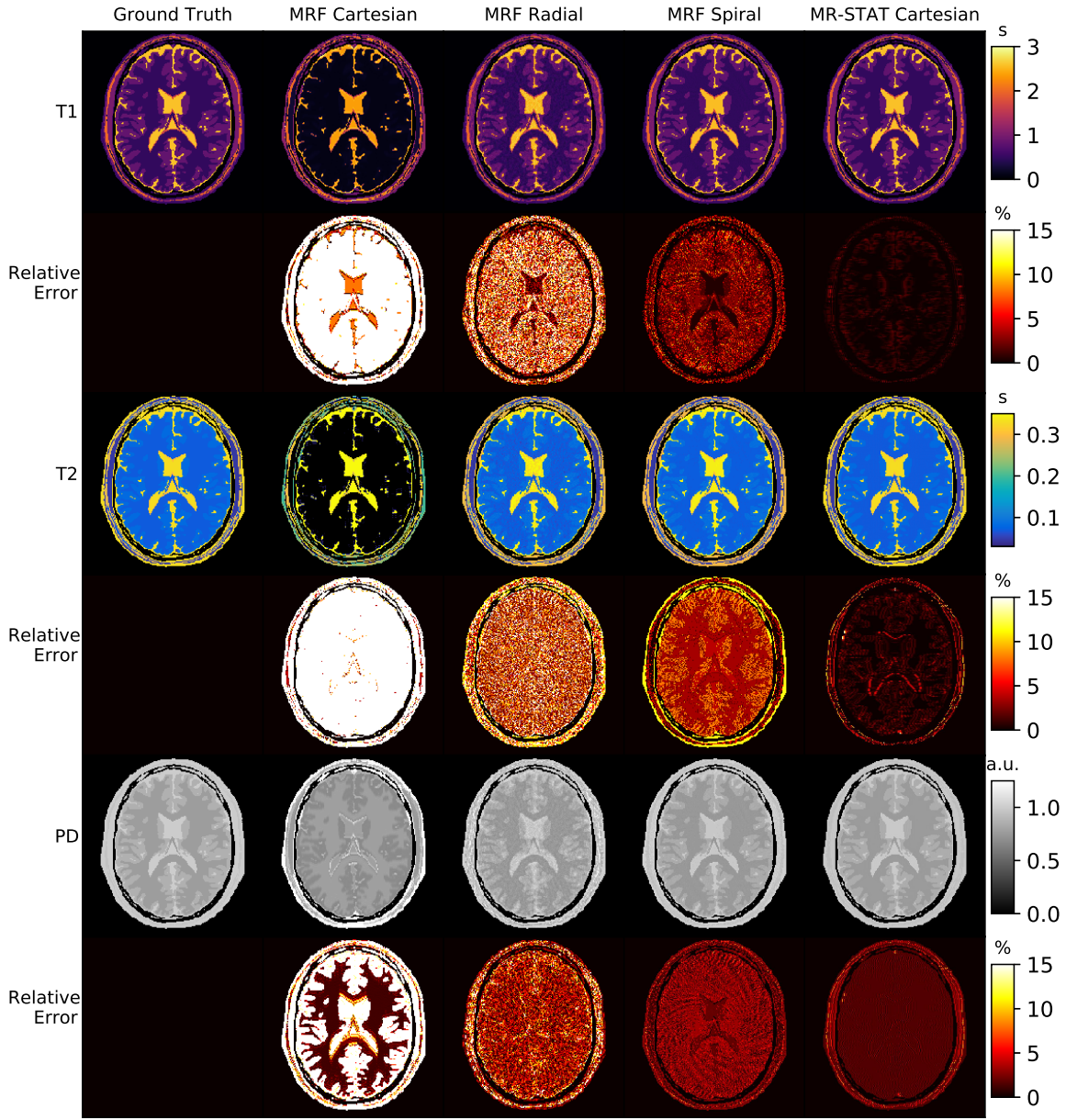}
    \caption{[First column] Ground truth $T_1, T_2$ and proton density maps for the numerical brain phantom. [Second, third and fourth columns] Reconstructed parameter maps and relative error maps for MRF with linear Cartesian, golden angle radial and golden angle spiral trajectories, respectively. [Fifth column] Reconstructed parameter maps and relative error maps for MR-STAT using a linear, Cartesian sampling trajectory. The MRF spiral and MR-STAT reconstructions both show excellent agreement with the ground truth values. The radial MRF reconstructions show residual aliasing artefacts and the Cartesian MRF reconstruction is heavily biased.}
    \label{fig:mrstat_vs_mrf}
\end{figure}

The $T_1, T_2$ and proton density maps reconstructed using (Cartesian) MR-STAT and Cartesian, radial and spiral MRF are shown in Figure \ref{fig:mrstat_vs_mrf} as well as the corresponding absolute relative error maps. It can be seen that the parameter maps reconstructed with either MR-STAT or spiral MRF are in excellent agreement with the ground truth. The radial MRF reconstructions show stronger residual streaking artefacts but in general the the estimated parameter values are close to the ground truth. For the Cartesian case the MRF reconstruction is unable to cope with the high undersampling (factor 192), resulting in severaly biased parameter maps. 

To quantify the quality of the reconstructions, normalized root mean square errors (``NRMSE"), high-frequency error norms (``HFEN", \cite{Ravishankar2011}, with standard deviation of 1.5 pixels) and mean absolute relative errors (``MAPE") were computed and are reported in Table \ref{tab:metrics}. It can be seen that the MR-STAT reconstruction results in the lowest NRSME and MAPE for all three parameters. The HFEN for the radial and spiral MRF and Cartesian MR-STAT reconstructions are similar.

\begin{figure}
    \centering
    \includegraphics[width=.99\textwidth]{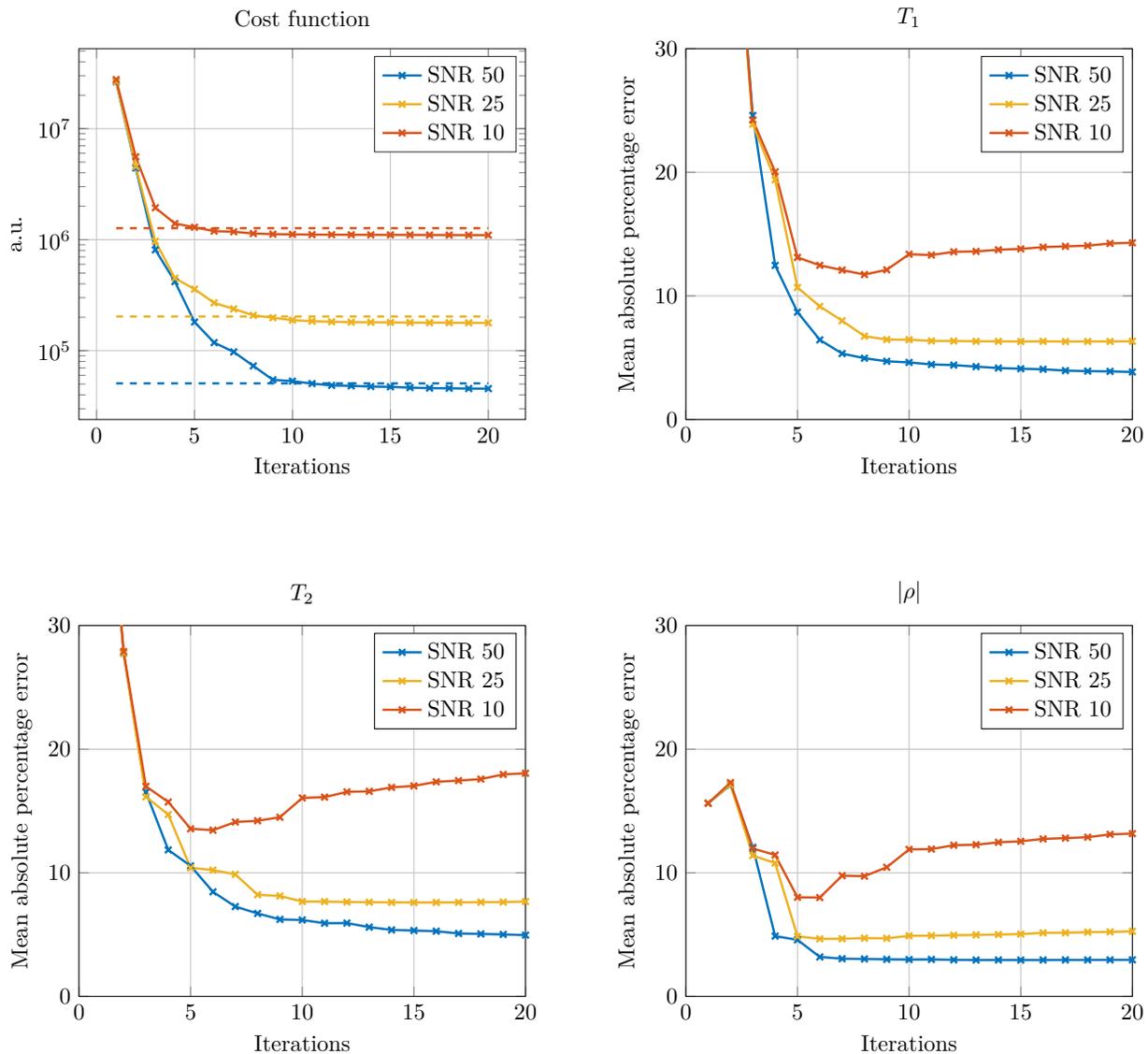}
    \caption{[Top Left] Convergence curves for the inexact Gauss-Newton MR-STAT method applied to data generated from the numerical brain phantom with different noise levels (SNR 50, 25, 10). In all cases the value of the cost function converges to the value expected based on the noise level. [Top right and bottom row] Mean absolute percentage errors for $T_1, T_2$ and proton density (magnitude) maps per iteration of the inexact Gauss-Newton method for different noise levels.}
    \label{fig:simulation_convergence}
\end{figure}

In Figure \ref{fig:simulation_convergence} convergence curves for MR-STAT with the inexact Gauss-Newton method for different SNR levels (50, 25 and 10) are shown as well as mean absolute percentage errors per iteration for $T_1$, $T_2$ and proton density. For the higher SNR cases the error values stabilize and the method converges after relatively few, e.g. ten, outer iterations. It can be seen that for the lowest SNR case, overfitting occurs after around six iterations. Based on these observations the number of outer iterations for the in-vivo case was chosen to be ten.

\subsection*{Gel Phantoms}

\begin{figure}
    \centering
    \includegraphics[width=.99\textwidth]{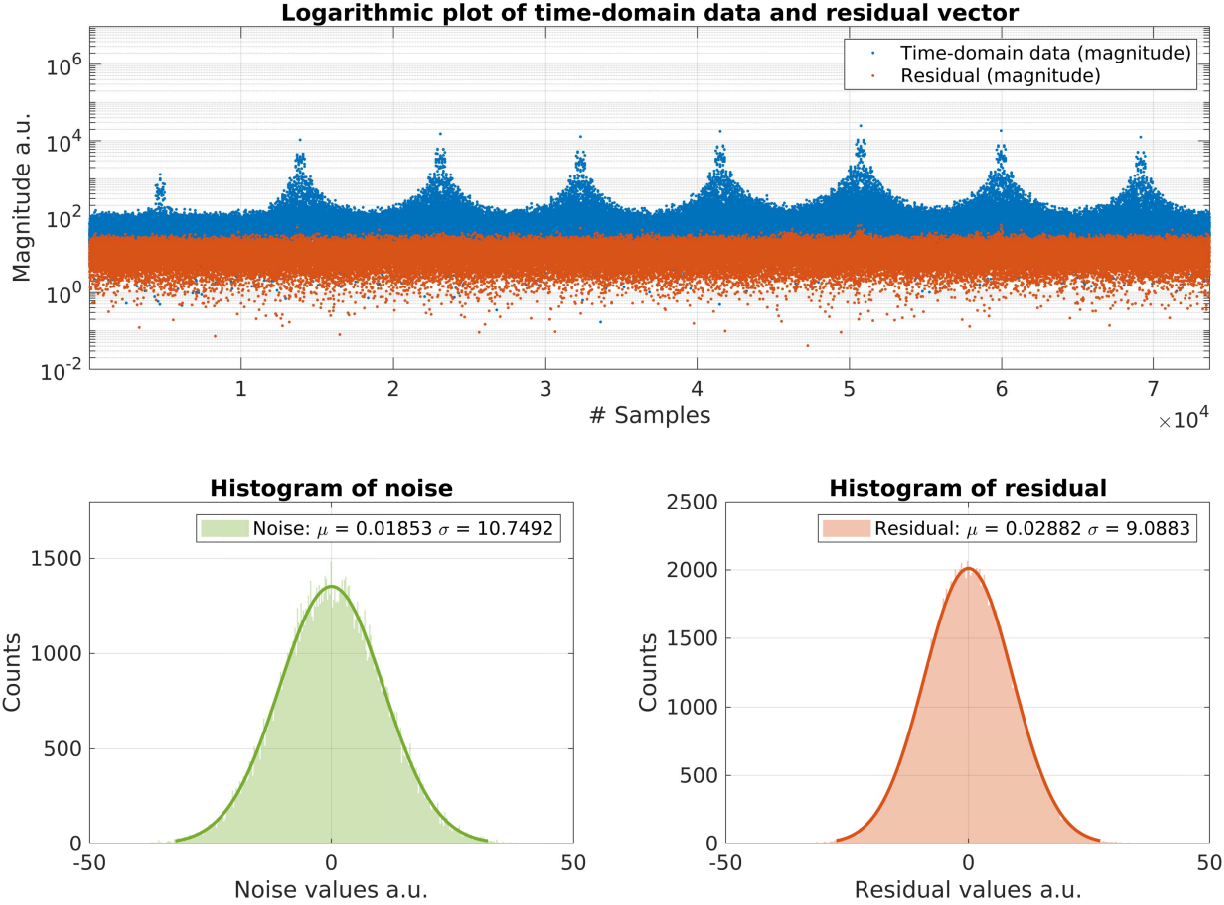}
    \caption{[Top row] Logarithmic plot of the magnitude of the measured time domain data obtained from the gel phantoms and the magnitude of the residual vector entries after the fifth iteration of the inexact Gauss-Newton method. [Bottom left] Histogram of noise values (real and imaginary values concatenated). The noise was measured using the receive channels right before the actual acquisition and it was subjected to the same pre-processing steps as the data used in the reconstruction (e.g. compression to a single channel using SVD). [Bottom right] Histogram of the residual vector entries (real and imaginary values concatenated) after the fifth iteration of the inexact Gauss-Newton method.}\label{fig:phantoms_maps}
\end{figure}

In Figure \ref{fig:phantoms_maps}, reconstructed $T_1$ and $T_2$ maps for the gel phantoms are shown and the mean $T_1$ and $T_2$ values per tube are compared to the gold standard measurements. It can be seen that the mean values are in excellent agreement. The mean values reported for the different repetitions of the MR-STAT scans are also in good agreement with each other (i.e. within standard deviations). In general the standard deviations for the reconstructed $T_2$ values is higher than for $T_1$ values, indicating a much stronger encoding of $T_1$ information into the signal which can be explained by the inversion pulse at the start of the sequence.

\begin{figure}
    \centering
    \includegraphics[width=.99\textwidth]{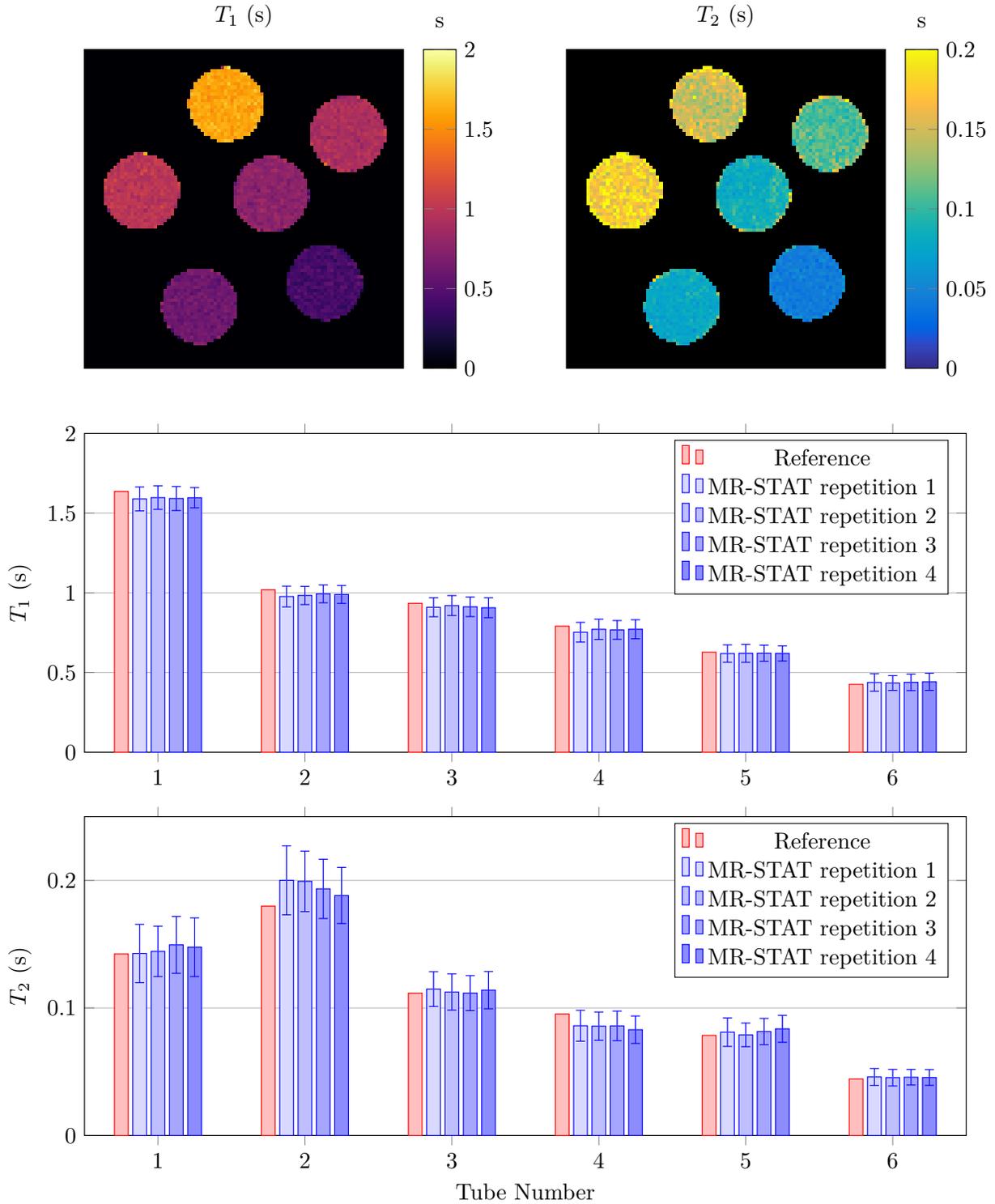}
    \caption{[Top row] $T_1$ and $T_2$ maps reconstructed with MR-STAT from gel-phantom data. [Middle and bottom rows] Comparison of mean $T_1$ and $T_2$ values obtained with MR-STAT and gold standard methods for each of the six gel phantom tubes. For MR-STAT the acquisition has been repeated four times. }\label{fig:phantoms_residual}
\end{figure}

To reconstruct the parameter maps, only five iterations of the reconstruction algorithm were needed and the total reconstruction time was approximately nine minutes using 32 computing cores. In Figure \ref{fig:phantoms_residual} a logarithmic plot of the measured signal magnitude and the residual vector after the fifth iteration are displayed for one of the MR-STAT repetitions. Histograms of the measured noise and the residual vectors are also shown. It can be seen that the residual vector follows a zero-mean Gaussian distribution with standard deviation similar to the noise, indicating that the model used in MR-STAT is able to adequately describe the measured time-domain signal.

\subsection*{High-resolution 2D brain scan}
\begin{figure}
    \centering
    \includegraphics[width=0.99\linewidth]{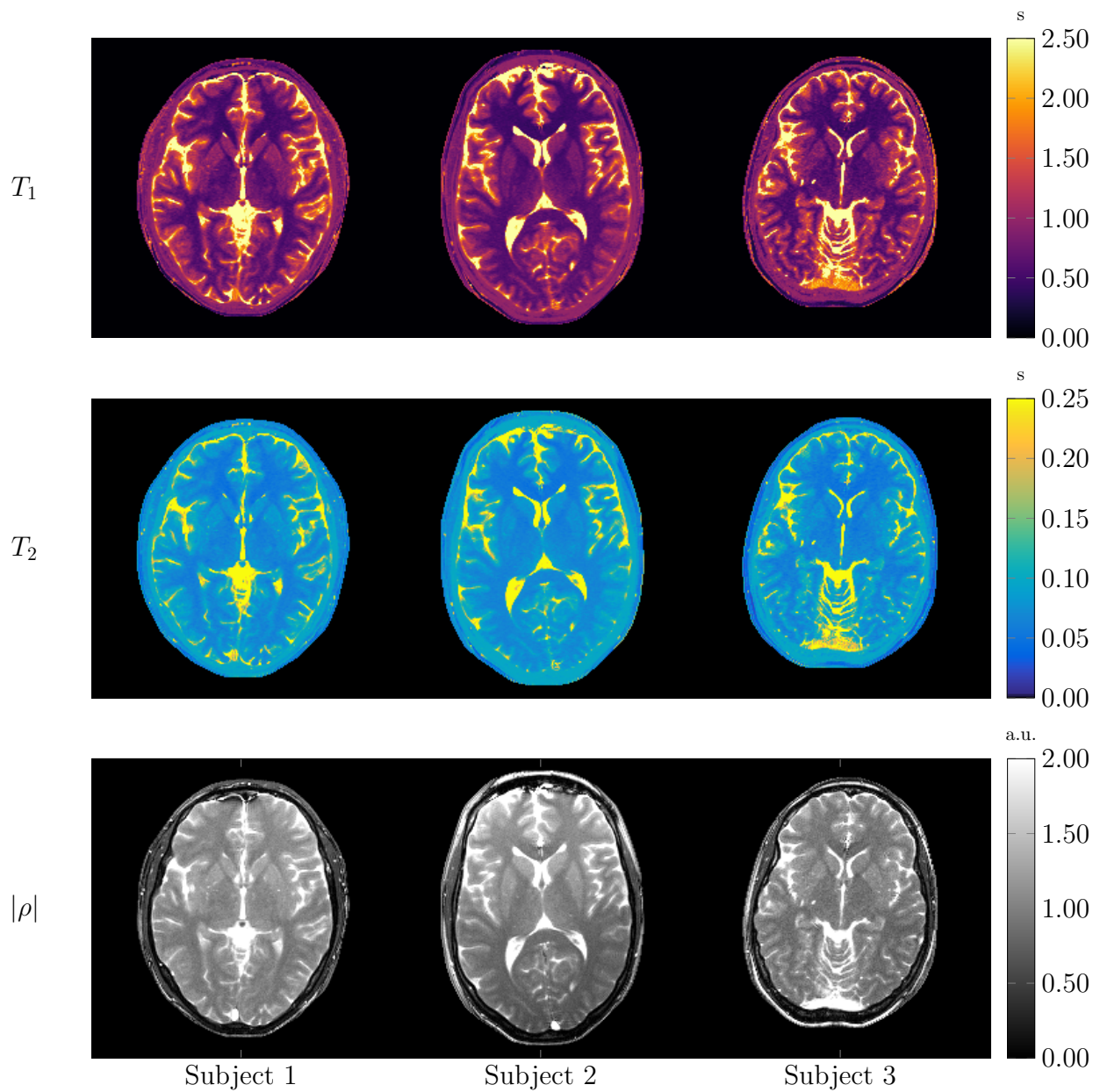}
    \caption{$T_1,T_2$ and proton density (magnitude) maps reconstructed with MR-STAT from in-vivo brain data obtained at 1.5 T (Philips, Ingenia) from multiple healthy volunteers. The in-plane resolution was $1 \times 1$ mm\textsuperscript{2} for all three subjects. For subjects $1$ and $2$ the acquisition time was $13.6$ s and the slice thickness was $5$ mm. For subject $3$ the acquisition time was $14.15$ s and the slice thickness was $3$ mm.}
    \label{fig:invivo_maps}
\end{figure}

In Figure \ref{fig:invivo_maps}, the reconstructed $T_1, T_2$ and proton density (magnitude) maps for the in-vivo brain scans performed on the three volunteers are shown. The maps show clear contrast between white matter, gray matter and cerebrospinal fluid (``CSF"). The maps corresponding to subject 3 appear noisier compared to the maps corresponding to subjects 1 and 2, which can be explained by the differences in slice thickness in the acquisition ($3$ mm vs $5$ mm). Mean $T_1$ and $T_2$ values and standard deviations in regions of white- and gray matter are reported in Table \ref{tab:meanvals}. The mean values are generally in good agreement with values found in literature for 1.5 T experiments \cite{Wright2008, Deoni2005, Breger1989} although we do observe an underestimation compared to some other studies, especially in white matter. We expect the underestimation is related to magnetization transfer that is known to affect the signal of balanced gradient-echo sequences (in a way that depends on the used $TR$ and RF pulse duration) \cite{Bieri2008,Rui2019}. The reconstruction time for each slice was approximately five hours using 64 cores.

\begin{figure}
    \centering
    \includegraphics[width=.99\textwidth]{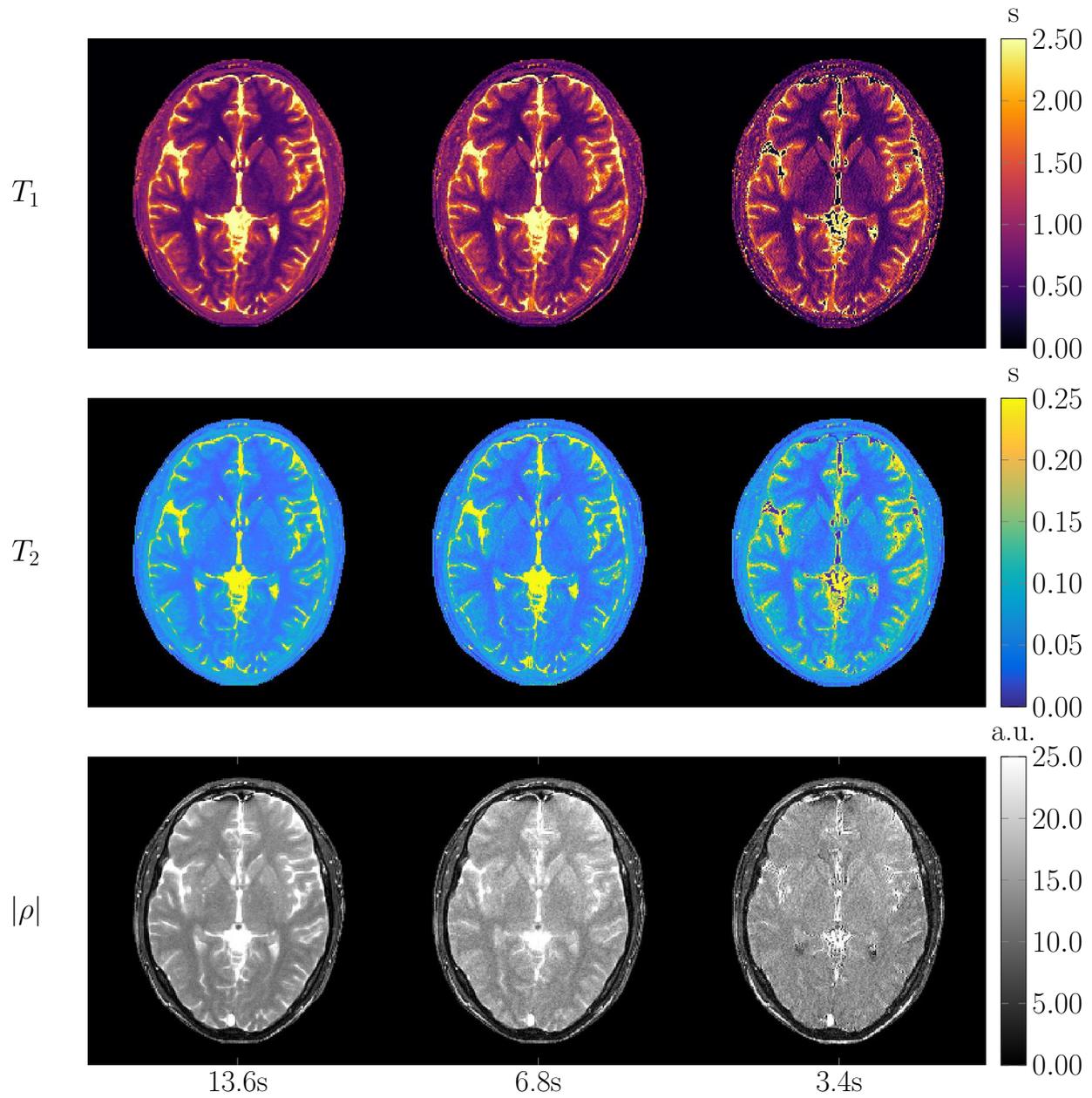}
    \caption{In-vivo $T_1,T_2$ and proton density (magnitude) maps at $1mm \times 1mm$ in-plane resolution reconstructed with MR-STAT based on acquisitions of, respectively, $13.6$ s, $6.8$ s and $3.4$ s on a 1.5 T MR system (Philips, Ingenia).}\label{fig:brain_maps_vs_kspaces}
\end{figure}

In Figure \ref{fig:brain_maps_vs_kspaces} we show $T_1, T_2$ and proton density (magnitude) for the same 2D brain slice but reconstructed using data corresponding to, respectively, $13.6$ s, $7.8$ s and $3.4$ s acquisitions. It can be seen that the maps corresponding to the 6.8 s acquisition are comparable to the maps corresponding to the $13.6$ s acquisition except that more noise is present. Depending on the application it might be more beneficial to repeat such a shorter sequence multiple times for noise averaging instead of scanning with the longer sequence. An added benefit of a shorter sequence duration is that the Bloch simulations are faster and thus reconstruction times are reduced by approximately the same factor with which the scantime is reduced. For the $3.4$ s acquisition the MR-STAT problem (Eq. \ref{eq:2}) is underdetermined in the sense that the number of datapoints is less than the number of unknowns in the problem. As can be seen in the reconstructed maps, this mostly results in biases in the CSF values. Note that in none of the reconstructions parallel imaging or compressed sensing techniques were utilized. 

\section*{Discussion \& Conclusion}
\label{sec:discussion}
MR-STAT is a framework for obtaining multiple quantitative parameter maps by fitting directly to measured time-domain data obtained from one short scan. Rather than relying on the FFT for spatial localisation of signal in a separate step, the spatial localisation and parameter estimation are performed simultaneously by iteratively solving a single non-linear optimization problem using a signal model that explicitly includes the spatial encoding gradients. The inherent large scale of the problem brings along new challenges in terms of computer memory requirements and computation times that make it difficult to perform MR-STAT reconstructions at high resolutions. To address these issues, we have presented a parallel and matrix-free reconstruction algorithm for MR-STAT and demonstrated that it can be used to generate high-resolution quantitative parameter maps.

All MR-STAT experiments in the current work have been performed with linear, Cartesian sampling strategies. This sampling strategy offers important advantages in the form of robustness to hardware imperfections (e.g. eddy currents, especially for gradient-balanced sequences \cite{Bieri2005,Bruijnen2019}), less susceptibility to $\Delta B_0$ related blurring artefacts \cite{Bernstein2004} and direct availability on clinical MR systems. Within the conventional MRF framework it is more challenging to work with Cartesian sampling strategies, as demonstrated using the simulation experiments. Studies that perform Cartesian MRF \cite{Koolstra2018,Buonincontri2016} therefore typically acquire multiple readout lines per snapshot, resulting in much longer acquisition times compared to non-Cartesian MRF acquisitions. A formal explanation of why Cartesian acquisitions are less suitable for MRF is reported in Stolk et al. \cite{Stolk2018}. More advanced iterative MRF reconstructions \cite{Zhao2016, Zhao2018, Asslander2018, Davies2014} might perform better with Cartesian sampling than the currently used MRF reconstructions (low-rank inversion followed by low-rank dictionary matching) and an in-depth comparison will be the subject of further studies. It should also be noted that neither the MR-STAT framework nor the currently proposed reconstruction algorithm are restricted to Cartesian sampling and further research is also aimed at incorporating non-Cartesian trajectories into MR-STAT.

An additional benefit of the volumetric signal model used in MR-STAT over FFT-based methods is that dynamic behaviour during the readouts (e.g. $T_2$-decay and $\Delta B_0$ induced phase accumulation) is taken into account. This may especially be beneficial for improving reconstructions based on acquisitions with long readouts (e.g. spiral readouts).

MR-STAT reconstructions are performed by solving a non-linear optimization problem using gradient-based iterative methods. No pre-computed dictionary is used. Compared to a dictionary-matching approaches there are no discretization errors and the reconstruction procedure is also flexible with respect to changes in sequence parameters (e.g. no rebuilding of a dictionary required when scan settings change). A downside of using iterative reconstruction algorithms to solve non-linear optimization problems is the risk of landing in a local minimum. In practice, with the currently used pulse sequence with smoothly changing flip angles and initial guess of the parameters, we have not encountered issues with local minima \cite{Sbrizzi2017}. 

Whereas with MRF the addition of new parameters results in an exponential increase in dictionary size (and thus also an exponential increase in dictionary generation and matching time), with MR-STAT additional parameters can be added at a quadratic increase in computation time. The quadratic increase can be explained as follows. The total number of parameters to be reconstructed $N$ increases linearly with the number of parameters per voxel ($N=N_pN_v$). Since the minimum number of time points $N_t$ that needs to be acquired - and thus simulated - is in the order of $N$, the computation time per Bloch simulation increases linearly as well. In addition, the number of partial derivative computations that needs to be performed per voxel also increases linearly. That is, both the number of rows and columns of the Jacobian $\mathbf{J}$ increase linearly, resulting in approximately a quadratic increase in computation time. In this respect we do note that, although currently $\Delta B_0$ maps are not reconstructed (because the employed bSSFP sequence used in this work is designed not to be sensitive to $\Delta B_0$ within the ``passband"), it is part of our all Bloch simulations and partial derivative computations. In addition, for the MR-STAT experiments described in the manuscript we used pulse sequences such that $N_t \approx 2N$ so that the problem remains overdetermined when an additional parameter is reconstructed. Therefore, assuming a pulse sequence is used that has sufficient $\Delta B_0$ encoding \cite{Shcherbakova2018,Wang2019}, we do not expect to see an increase in computation times when reconstructing $\Delta B_0$ as an additional parameter. 

For the phantom experiment we observed that the noise levels was reached for the residuals. However, this was not observed for the in-vivo case as certain effects are still accounted for in the model. Examples include patient motion, blood flow, magnetization transfer and diffusion effects.

A limitation of the proposed method is that at this moment reconstruction times are still long for high-resolution scans, especially when compared to the dictionary matching procedures used in MRF. Even when employing a high performance computing cluster, reconstruction times in the order of hours for a single 2D brain slice. Although possible from a memory point-of-view, 3D reconstructions will take too long for practical purposes with the current reconstruction setup. The main bottleneck in the reconstructions is formed by the partial derivative computations needed to solve equation \eqref{eq:4}. Further research is aimed at performing these computations on GPU architectures \cite{Xanthis2014,Kose2017}, reducing the computational effort through algorithmic improvements \cite{Heide2019} and through the use of surrogate models \cite{Yang2018}. Together with (cloud) computing resources becoming cheaper and more accessible over time, we believe it is possible to accelerate the computations to an extent that MR-STAT becomes applicable in clinical settings.

Further research is also aimed at reduction of acquisiton time and improving precision and accuracy of the MR-STAT parameter maps by incorporating parallel imaging \cite{Heide2019_2}, compressed sensing and through sequence optimization. 

The main aim of the MR-STAT project is to explore possibilities to achieve very short acquisition times beyond what is possible with FFT-based frameworks. Although the MR-STAT framework in principle allows for much flexibility in the data acquisition process (e.g. non-Cartesian acquisitions), in the current work we have opted for Cartesian sampling patterns because of their robustness to hardware imperfections and because they clearly exemplify the benefits of skipping the FFT step (i.e. no introduction of artificial aliasing noise through application of the FFT on undersampled k-spaces). An additional benefit is the direct availability of such sequences on clinical MR systems. In the current work we used constantly varying flip angle trains, however, as shown in Supporting Information S3, MR-STAT could even be used with Cartesian bSSFP sequences with a fixed flip angle per k-space that require little to no pulse programming for their implementation. 

\section*{Acknowledgements}
This work was funded by the Dutch Technology Foundation, grant \#14125.

\bibliographystyle{unsrt}
\bibliography{../bibtex/main}
% \input{main.bbl}
  
%%TC:ignore
\clearpage
\listoffigures 
\clearpage

\clearpage

\clearpage

\clearpage

\clearpage
%%TC:endignore

\clearpage
\listoftables 
\clearpage

\begin{table}[bt]
\centering
\caption{On-disk sizes of MR-STAT matrices for $N_p=4$ and $N_t = N = 4 \times N_v$ for an increasing number of voxels $N_v$. The memory sizes are computed as $2 \cdot N_v \cdot N_t \cdot 8$ bytes ($\mathbf{M}$), $2 \cdot N \cdot N_t \cdot 8$ bytes ($\mathbf{J}$) and $N^2 \cdot 8$ bytes ($\mathbf{H}$) respectively. The factors of 2 come from the real and imaginary components and the factor of 8 represents the bytes necessary to store 64-bit floating point numbers.}\label{tab:matrixsizes}
% \begin{threeparttable}
\begin{tabular}{|l|l|l|l|l|}
% \headrow
\hline
\thead{Image size} & \thead{Voxels $(N_v)$} & \thead{$\mathbf{M}$} & \thead{$\mathbf{J}$} & \thead{$\mathbf{H}$}\\ \hline
    $\,\,\,64  \times 64$   & $4.096$       & 1 GB  & 4  GB & 2  GB  \\
    $128 \times 128$        & $16.384$      & 16 GB   & 64  GB  & 32  GB   \\
    $256 \times 256$        & $65.536$      & 256 GB  & 1.024 GB  & 512 GB   \\
    $512 \times 512$        & $262.144$     & 4.096 GB  & 16.384 GB &  8.192 GB  \\
\hline  % Please only put a hline at the end of the table
\end{tabular}
% \end{threeparttable}
\end{table}

\begin{table}[bt]
    \centering
    \caption{Acquisition parameters for in-vivo MR-STAT brain scans.}\label{tab:acqpars}
    % \begin{threeparttable}
    \begin{tabular}{|l|l|l|}
    % \headrow
    \hline
    \thead{Acquisition parameter} & \thead{Subjects 1 and 2 } & \thead{Subject 3 } \\ \hline
        Field strength      & $1.5$ T            & $1.5$ T            \\ 
        In-plane resolution & $1$ mm $\times$ $1$ mm & $1$ mm $\times$ $1$ mm  \\ 
        Field-of-view       & $224$ mm $\times$ $224$ mm & $224$ mm $\times$ $224$ mm \\  
        Slice thickness     & $5$ mm             & $3$ mm             \\ 
        TR                  & $7.6$ ms           & $7.9$ ms           \\ 
        TE                  & $3.8$ ms           & $3.95$ ms          \\ 
        Readout bandwidth   & $85.6$ kHz         & $85.6$ kHz         \\ 
        Pulse duration      & $0.76$ ms         & $0.81$ ms         \\ 
        Scan time           & $13.6$ s           & $14.15$ s               \\
    \hline  % Please only put a hline at the end of the table
    \end{tabular}
    % \end{threeparttable}
\end{table}
\clearpage

\begin{table}[bt]
    \centering
    \caption{Three different error metrics (NRMSE, HFEN, MAPE) computed for the MRF (Cartesian, Radial and Spiral) and MR-STAT (Cartesian) reconstructions on the numerical brain phantom. No noise was added to the data in these reconstructions. The MR-STAT reconstructions result in the lowest errors because the reconstructions do not suffer from undersampling artefacts and because there are no discretization errors due to a finite dictionary.}\label{tab:metrics}
    % \begin{threeparttable}
    \begin{tabular}{|l|l|l|l|l|l|l|}
    % \headrow
    \hline
    Parameter & Metric & Units & MRF Cartesian & MRF Radial & MRF Spiral & MR-STAT Cartesian \\ \hline
        $T_1$ & NRMSE & [a.u.] & 0.2302 & 0.0432 & 0.0110 & 0.0025 \\
              & MAPE  & [\%]   & 65.6   & 8.5    & 2.5    & 0.4 \\
              & HFEN  & [a.u.] & 18.1   & 15.6   & 15.7   & 15.7 \\
        $T_2$ & NRMSE & [a.u.] & 0.2486 & 0.0756 & 0.0492 & 0.0048 \\
              & MAPE  & [\%]   & 56.1   & 8.1    & 4.9    & 0.9 \\
              & HFEN  & [a.u.] & 3.2    & 2.5    & 2.4    & 2.5 \\
        $\rho$& NRMSE & [a.u.] & 1.0979 & 0.2626 & 0.1129 & 0.0830 \\
              & MAPE  & [\%]   & 15.8   & 4.8    & 2.7    & 1.8 \\
              & HFEN  & [a.u.] & 8.5   & 5.6     & 5.5    & 5.4 \\
    \hline
    \end{tabular}
    % \end{threeparttable}
    \end{table}

\begin{table}[bt]
\centering
\caption{Mean $T_1$ and $T_2$ values and standard deviation in white- and gray matter regions for each of the three in-vivo brain scans.}\label{tab:meanvals}
% \begin{threeparttable}
\begin{tabular}{|l|l|c|c|}
% \headrow
\hline
\thead{Tissue type} & \thead{Subject} & \thead{$T_1$} & \thead{$T_2$}\\ \hline
            Frontal white matter    & 1 & $505 \pm 48$ ms & $53.3 \pm 4.0$ ms \\
                                    & 2 & $542 \pm 48$ ms & $57.4 \pm 3.8$ ms \\
                                    & 3 & $519 \pm 54$ ms & $56.1 \pm 4.3$ ms \\
            Putamen (gray matter)   & 1 & $874 \pm 64$ ms & $74.8 \pm 4.4$ ms \\
                                    & 2 & $956 \pm 66$ ms & $80.2 \pm 4.5$ ms \\
                                    & 3 & $895 \pm 107$ ms & $78.4 \pm 7.0$ ms \\
\hline  % Please only put a hline at the end of the table
\end{tabular}
% \end{threeparttable}
\end{table}

\clearpage
\listofalgorithms 
\clearpage

\begin{algorithm}
    \caption{Minimize $f(\boldsymbol{\alpha}) = \frac{1}{2}\|\mathbf{d} - \mathbf{s(\boldsymbol{\alpha})} \|_2^2$}
    \begin{algorithmic}[0]
        \REQUIRE Initial guess $\boldsymbol{\alpha}$
        \WHILE{not converged}
        \STATE 1. Compute residual: $\mathbf{r} = \mathbf{d} - \mathbf{s} = \mathbf{d} - \mathbf{M}\boldsymbol{\rho}$
        \STATE 2. Compute gradient: $\mathbf{g} = \Re \left( \mathbf{J}^H \mathbf{r} \right)$
        \STATE 3. Solve linear system: $\mathbf{H} \mathbf{p} = -\mathbf{g}$
        \STATE 4. Update parameters: $\boldsymbol{\alpha} = \boldsymbol{\alpha} + \mathbf{p}$
        \ENDWHILE
    \end{algorithmic}
    \label{alg:alg1}
\end{algorithm}

\begin{algorithm}
    \caption{Compute $\mathbf{s}(\boldsymbol{\alpha}) = \mathbf{M}\boldsymbol{\rho}$ (matrix-free, serial)}
    \begin{algorithmic}[0]
        \STATE Initialize $\mathbf{s}$ = zeros($N_t$,1)
        \FOR{j $\leftarrow$ $1$ to $N_v$}
        \STATE 1. Integrate Bloch equations in time to obtain $\mathbf{m}_j = \left[ m_j(t_1), \ldots, m_j(t_{N_t})\right]^T$
        \STATE 2. Set $\mathbf{s} = \mathbf{s} + \rho_j \mathbf{m}_j$
        \ENDFOR
        \STATE Return $\mathbf{s}$
    \end{algorithmic}
    \label{alg:algMv}
\end{algorithm}

\begin{algorithm}
    \caption{Compute $\mathbf{s}(\boldsymbol{\alpha}) = \mathbf{M}\boldsymbol{\rho}$ (matrix-free, parallel)}
    \begin{algorithmic}[0]
        \REQUIRE Master process $p_m$, slave processes $p_i$ for $i \in [1, \ldots, N_c]$.
        % \STATE Initialize $\mathbf{s}$ = zeros(Ns,1)
        \STATE 1. $p_m$ distributes $\boldsymbol{\alpha}$: each $p_i$ receives the parameters associated with voxels $ [(i-1) * N_v + 1, \ldots, i*N_v]$.
        \STATE 2. Each $p_i$ uses Algorithm \ref{alg:algMv} to compute a  ``local'' version of the signal $\tilde{\mathbf{s}}_{i}$.
        \STATE 3. Each $p_i$ communicates $\tilde{\mathbf{s}}_{i}$ back to the $p_m$.
        \STATE 4. On $p_m$ the signal $\mathbf{s}$ is computed as $\mathbf{s} = \sum_{i=1}^{N_c} \tilde{\mathbf{s}}_{i}$.
    \end{algorithmic}
    \label{alg:algMv_par}
\end{algorithm}

\begin{algorithm}
    \caption{Compute $\mathbf{g}(\boldsymbol{\alpha}) = \Re(\mathbf{J}^H \mathbf{r})$ (matrix-free, serial)}
    \begin{algorithmic}[0]
        \STATE Initialize $\mathbf{g}$ = zeros($N$,1)
        \FOR{j $\leftarrow$ 1 to $N_v$}
        \FOR{k $\leftarrow$ 1 to $N_p$}
        \STATE 1. Set $i = j + (k-1)N_v$
        \STATE 2. Compute partial derivative $\frac{\partial \mathbf{r}}{\partial \alpha_{i}}  = -\frac{\partial \left(\rho_j \mathbf{m}_{j} \right)}{\partial \alpha_{i}}$
        \STATE 3. Set $\mathbf{g}[i] = \Re \left( \left<\frac{\partial \mathbf{r}}{\partial \alpha_{i}}, \mathbf{r} \right> \right)$
        \ENDFOR
        \ENDFOR
        \STATE Return $\mathbf{g}$
    \end{algorithmic}
    \label{alg:algJtv}
\end{algorithm}

\begin{algorithm}
    \caption{Compute $\mathbf{g}(\boldsymbol{\alpha}) = \Re (\mathbf{J}^H \mathbf{r})$ (matrix-free, parallel)}
    \begin{algorithmic}[0]
        \REQUIRE Master process $p_M$, slave processes $p_i$ for $i \in [1, \ldots, N_c]$.
        \STATE 1. $p_m$ distributes $\boldsymbol{\alpha}$: each $p_i$ receives the parameters associated with voxels $ [(i-1) * N_v + 1, \ldots, i*N_v]$.
        \STATE 2. $p_m$ distributes $\mathbf{r}$ to each $p_i$.
        \STATE 3. Each $p_i$ uses Algorithm \ref{alg:algJtv} to compute a  ``local'' gradient $\tilde{\mathbf{g}}_{i}$.
        \STATE 4. Each $p_i$ communicates $\tilde{\mathbf{g}}_{i}$ back to the $p_m$.
        \STATE 5. On $p_m$ the gradient $\mathbf{s}$ is computed by vertical concatenation of the $\tilde{\mathbf{g}}_{i}$.
    \end{algorithmic}
    \label{alg:algJtv_par}
\end{algorithm}

\begin{algorithm}
    \caption{(Inexact Gauss-Newton MR-STAT) Minimize $f(\boldsymbol{\alpha}) = \frac{1}{2}\|\mathbf{d} - \mathbf{s(\boldsymbol{\alpha})} \|_2^2$}
    \begin{algorithmic}
        \REQUIRE Initial guess $\boldsymbol{\alpha_0}$, initial trust radius $\Delta_0$
        \STATE
        \WHILE{not converged}
        \STATE Compute $\mathbf{s}$ (matrix-free, parallel)
        \STATE Set $\mathbf{r} = \mathbf{d} - \mathbf{s}$
        \STATE Compute $\mathbf{g} = \mathbf{J}^H \mathbf{r}$ (matrix-free, parallel)
        \STATE Solve with CG iterations (inner GN loop):
        \STATE $\quad \Re \left(\mathbf{J}^H \mathbf{J} \right) \mathbf{p} = -\mathbf{g}$ (matrix-free, parallel)
        \STATE Set $\boldsymbol{\alpha} = \boldsymbol{\alpha} + \mathbf{p}$
        \ENDWHILE
    \end{algorithmic}
    \label{alg:inexact_gn}
\end{algorithm}

%%TC:ignore
\clearpage
\section*{Legend of Supporting Information}
  \begin{enumerate}
      \item \textbf{Supporting Information S1:} MR-STAT pulse sequence
      \item \textbf{Supporting Information S2:} Algorithmic Differentiation of Bloch equation solver
      \item \textbf{Supporting Information S3:} MR-STAT with Cartesian sampling and constant flip angle per k-space
  \end{enumerate}
%%TC:endignore
\end{document}

\section*{High resolution in-vivo MR-STAT using a matrix-free and parallelized reconstruction algorithm}

\subsection*{Supporting Information S1: MR-STAT pulse sequence}

\begin{figure}[H]
    \centering
    \includegraphics[width=0.97\linewidth]{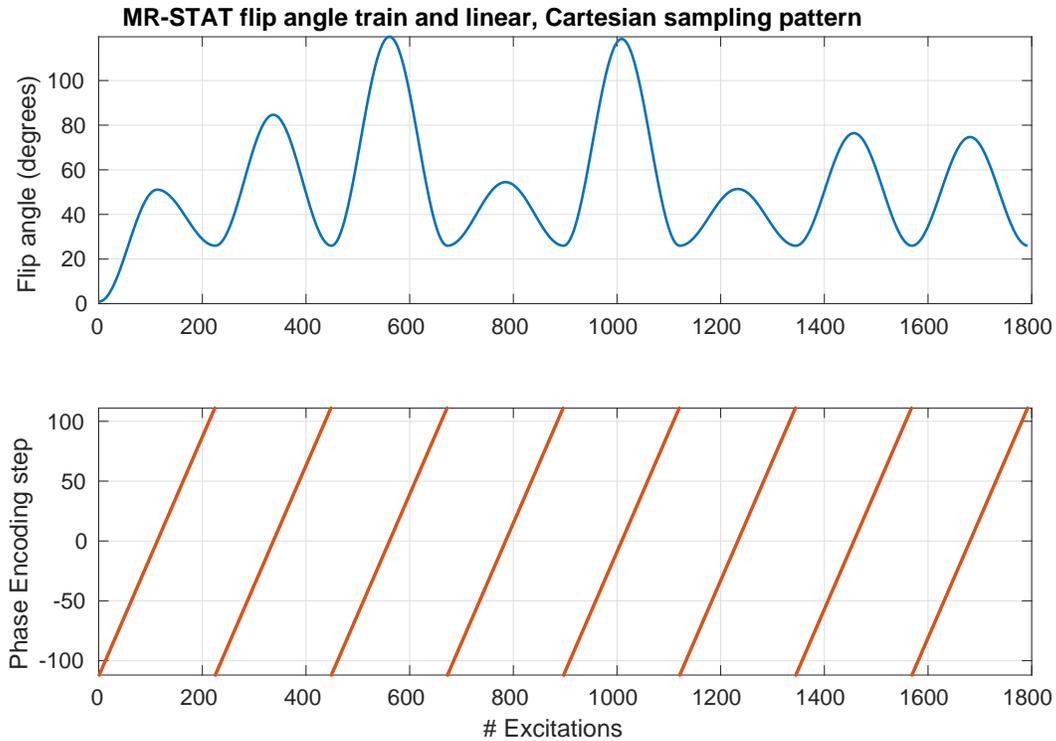}
    \caption{Overview of the flip angle train and the Cartesian gradient-encoding order used in the 2D (transient-state) MR-STAT pulse sequence.}
    \label{fig:mrstat_sequence}
\end{figure}

For the simulation, gel phantom and in-vivo MR-STAT experiments discussed in the main text a pulse sequence was used with linear, Cartesian sampling and flip angles that change each TR in accordance with a smoothly varying pattern shown in Supporting Information Figure \ref{fig:mrstat_sequence}. The flip angle pattern starts at zero and then consists of connected sine-squared waves with randomly generated peak amplitudes. The sine-squared waves are aligned such that the central line of k-space is sampled at the peaks.

\clearpage

\subsection*{Supporting Information S2: Algorithmic Differentiation of Bloch equation solver}
The Bloch equations define a system of (linear, inhomogeneous) ordinary differential equations with time-dependent coefficients (e.g. RF and gradient waveforms). To numerically integrate the Bloch equations, the time domain is discretized into time intervals during which the RF and gradient waveforms are assumed to be constant. Given the magnetization $\mathbf{m}_{t}=\left[m_{t,x},m_{t,y},m_{t,z}\right]^{T}$
in a voxel with coordinates $\mathbf{r}=\left[x,y,z\right]^{T}$ at
time $t$, to find (an approximation to) the magnetization at time
$t+\Delta t$ we first apply a rotation induced by the (complex) RF
pulse and gradients $\mathbf{GR}=(GR_{x},GR_{y},GR_{z})$ that are
present during the time interval $[t,t+\Delta t]$ to obtain the rotated
magnetization $\mathbf{m}_{rot}$. Afterwards we apply $T_{1}$ and $T_{2}$
induced decay and regrowth to $\mathbf{m}_{rot}$ to obtain $\mathbf{m}_{t+\Delta t}.$
We refer to \cite{Valenberg2015} for more details on the discretization.

To apply the rotation during each time interval, we first compute a rotation vector $\mathbf{a}$
as
\begin{equation}
\mathbf{a}=\gamma\Delta t\begin{bmatrix}-|B_{1}^{+}|\text{Re}(RF)\\
|B_{1}^{+}|\text{Im}(RF)\\
\mathbf{GR}\cdot\mathbf{r}+\Delta B_{0}/\gamma
\end{bmatrix}.
\end{equation}
Now define $\mathbf{k}=\mathbf{a}/\|\mathbf{a}\|_{2}$ (i.e. $\mathbf{k}$
is a unit vector pointing in the same direction as $\mathbf{a}$)
and $\theta=\|\mathbf{a}\|_{2}.$ By using Rodrigues' rotation formula
we can compute $\mathbf{m}_{rot}$ as:
\begin{equation}
\mathbf{m}_{rot}=\cos(\theta)\mathbf{m}_{t}+\sin(\theta)(\mathbf{k}\times\mathbf{m}_{t})+(1-\cos(\theta))(\mathbf{k}\cdot\mathbf{m}_{t})\mathbf{k}.\label{eq:m_rot}
\end{equation}

To apply the decay and regrowth we first compute $E_{1}:=\exp(-\Delta t/T_{1})$
and $E_{2}:=\exp(-\Delta t/T_{2})$. Then, the magnetization at time
$t+\Delta t$ is obtained as follows:
\begin{equation}
\mathbf{m}_{t+\Delta t}=\mathbf{m}_{rot}\otimes[E_{2},E_{2},E_{1}]^{T}+[0,0,1-E_{1}]^{T}.\label{eq:m_decay}
\end{equation}
The simulation is initialized with the magnetization in equilibrium
position along the $z-$axis, i.e. $\mathbf{m}_{0}=[0,0,1]^{T}.$ 

Given the update formula for the magnetization, we can apply algorithmic
differentiation to obtain update formulas for the partial derivatives.
We illustrate this procedure for partial derivatives with respect
to $T_{1}.$ First of all, note that from Eq. \ref{eq:m_rot} we can derive
that
\begin{eqnarray}
\frac{\partial\mathbf{m}_{rot}}{\partial T_{1}} & = & \frac{\partial}{\partial T_{1}}\left(\cos(\theta)\mathbf{m}_{t}\right)+\frac{\partial}{\partial T_{1}}\left(\sin(\theta)(\mathbf{k}\times\mathbf{m}_{t})\right)  \nonumber\\
& &+ \frac{\partial}{\partial T_{1}}\left((1-\cos(\theta))(\mathbf{k}\cdot\mathbf{m}_{t})\mathbf{k}\right).
\end{eqnarray}
Since neither the rotation axis $\mathbf{k}$ nor the rotation angle
$\theta$ depend on $T_{1},$ it follows that
\begin{eqnarray}
\frac{\partial\mathbf{m}_{rot}}{\partial T_{1}} & = & \cos(\theta)\frac{\partial\mathbf{m}_{t}}{\partial T_{1}}+\sin(\theta)\frac{\partial(\mathbf{k}\times\mathbf{m}_{t})}{\partial T_{1}} \nonumber \\ 
& &+(1-\cos(\theta))\frac{\partial(\mathbf{k}\cdot\mathbf{m}_{t})}{\partial T_{1}}\mathbf{k} \nonumber\\
 & = & \cos(\theta)\frac{\partial\mathbf{m}_{t}}{\partial T_{1}}+\sin(\theta)\left(\mathbf{k}\times\frac{\partial\mathbf{m}_{t}}{\partial T_{1}}\right) \nonumber \\
 & &+(1-\cos(\theta))\left(\mathbf{k}\cdot\frac{\partial\mathbf{m}_{t}}{\partial T_{1}}\right)\mathbf{k}.\label{eq:dm_rot}
\end{eqnarray}
Next, using Eq. \ref{eq:m_decay}, we find that
\begin{eqnarray}
\frac{\partial\mathbf{m}_{t+1}}{\partial T_{1}} & = & \frac{\partial\mathbf{m}_{rot}}{\partial T_{1}}\otimes[E_{2},E_{2},E_{1}]^{T}+\mathbf{m}_{rot}\otimes\frac{\partial[E_{2},E_{2},E_{1}]^{T}}{\partial T_{1}} \nonumber \\ 
& &+\frac{\partial[0,0,1-E_{1}]^{T}}{\partial T_{1}} \nonumber \\
 & = & \frac{\partial\mathbf{m}_{rot}}{\partial T_{1}}\otimes[E_{2},E_{2},E_{1}]^{T}+\mathbf{m}_{rot}\otimes[0,0,\frac{\Delta t}{T_{1}^{2}}E_{1}]^{T} \nonumber \\
 & &+[0,0,-\frac{\Delta t}{T_{1}^{2}}E_{1}]^{T}.\label{eq:dm_decay}
\end{eqnarray}
By combining Eq. \ref{eq:dm_rot} and Eq. \ref{eq:dm_decay} we have a recursive update formula for the partial derivatives. Since $\mathbf{m}_{0}=\left[0,0,1\right]^{T}$ is independent of $T_{1},$
the partial derivatives are initialized with $\frac{\partial\mathbf{m}_{0}}{\partial T_{1}}=\left[0,0,0\right]^{T}.$ The procedure to find partial derivatives with respect to the other parameters is similar. 

Although toolboxes exist that can automate the process of computing the derivatives from update, we have opted for a manual implementation to optimize for performance. Compared to the finite-difference method utilized in Sbrizzi et al \cite{SBRIZZI201856} algorithmic differentiation does not require the (non-trivial) choice of step sizes. And whereas the finite difference method approximates derivatives, with algorithmic differentiation the derivatives are exact. On top of that, our implementation of the algorithmic differentiation runs faster than the finite difference method by approximately a factor of two. This speedup can be partially explained by the fact that the relatively expensive $\sin, \cos$ (for rotations) and $\exp$ (for decay and regrowth) terms that need to be computed at each time step of the numerical integration can be efficiently reused for computing the partial derivatives.

\clearpage

\clearpage
\subsection*{Supporting Information S3: MR-STAT with Cartesian sampling and constant flip angle per k-space}

\begin{figure}
\includegraphics[width=1\linewidth]{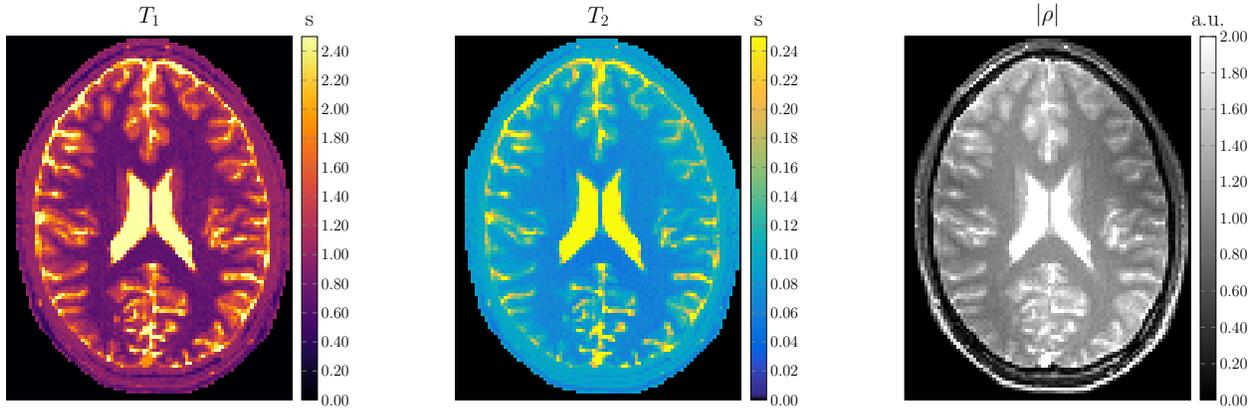}
\caption{Reconstructed $T_1,T_2$ and proton density (magnitude) maps for a 2D slice of an in-vivo brain of a healthy volunteer at $1.8$ mm $\times$ $1.8$ mm in-plane resolution using a the pulse sequence with constant flip angle per k-space on a 3 T (Philips, Ingenia) MR system.}\label{fig:constant_fa_maps}
\end{figure}

\begin{figure}
\centering
\includegraphics[width=1\linewidth]{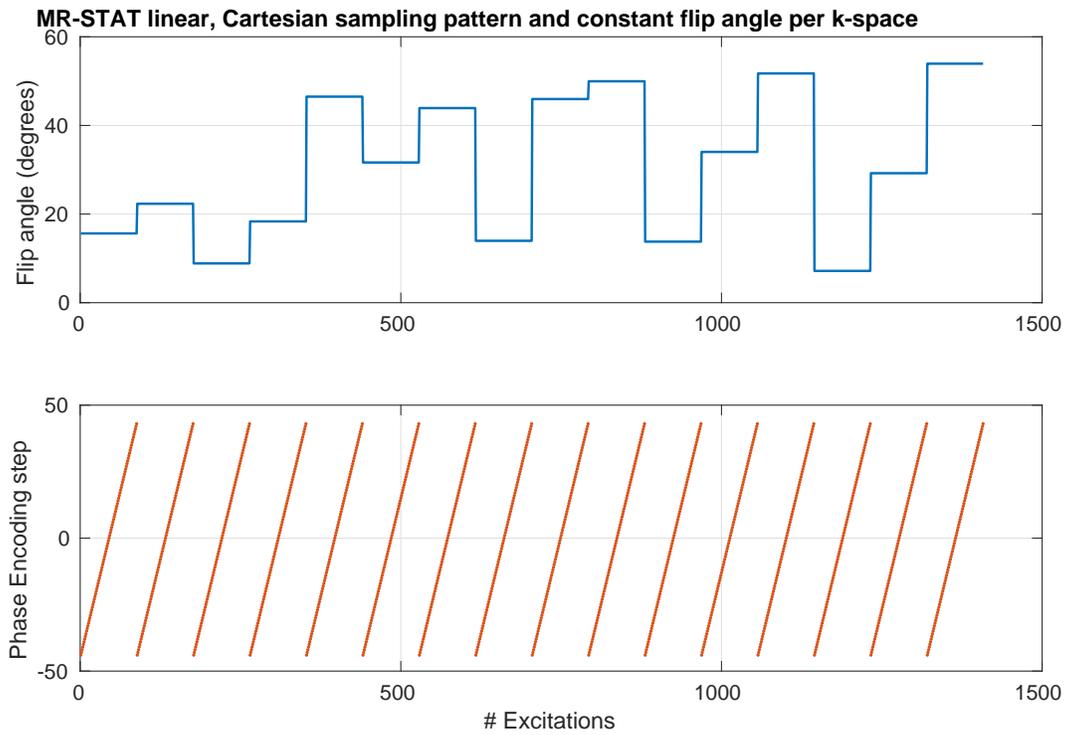}
\caption{Visualization of the pulse sequence with linear, Cartesian sampling and constant flip angle per k-space.}
\label{fig:constant_fa_seq} 
\end{figure}

To demonstrate the flexibility of the MR-STAT framework we also acquired in-vivo brain data using a pulse sequence that has a constant flip angles per k-space (but different flip angle between the k-spaces). The data was acquired on a 3 T clinical MR-system with the following acquisition parameters: in-plane resolution $1.8$ mm $\times$ $1.8$ mm, field-of-view $201.5$ mm $\times$ $158$ mm, TR $5.6$ ms, TE $2.8$ ms, slice thickness $3$ mm, readout bandwidth $77.2$ kHz, number of k-spaces $16$, total scan time $7.82$ s. The inexact Gauss-Newton method with ten outer iterations and fifteen inner iterations was used for the reconstruction. The total reconstruction time was approximately 30 minutes using 64 cores on the computing cluster. The reconstructed parameter maps and the flip angle train are shown in Supporting Information Figure \ref{fig:constant_fa_maps} and Supporting Information Figure \ref{fig:constant_fa_seq} respectively. The results demonstrate that excellent quality parameter maps can be obtained using in a short time using a pulse sequence that is readily available even on older clinical MR-systems and requires little to no pulse programming.

\clearpage

\clearpage
% \bibliographystyle{unsrt}
% \bibliography{main}